# Understanding Molecular Dynamics with Stochastic Processes *via* Real or Virtual Dynamics

Dezhang Li[1,*] , Zifei Chen[1,*], Zhijun Zhang[1], and Jian Liu[1, †]

[1.] Beijing National Laboratory for Molecular Sciences, Institute of Theoretical and Computational Chemistry, College of Chemistry and Molecular Engineering, Peking University, Beijing 100871, China

(Invited contribution to the CJCP Special Topic Issue on “Dynamics”)

† Electronic mail: jianliupku@pku.edu.cn

* The two authors contributed equally to the work

## Abstract

Molecular dynamics with the stochastic process provides a convenient way to compute structural and thermodynamic properties of chemical, biological, and materials systems. It is demonstrated that the virtual dynamics case that we proposed for the Langevin equation [J. Chem. Phys. 147, 184104 (2017)] in principle exists in other types of stochastic thermostats as well. The recommended "middle" scheme [J. Chem. Phys. 147, 034109 (2017)] of the Andersen thermostat is investigated as an example. As shown by both analytic and numerical results, while the real and virtual dynamics cases approach the same plateau of the characteristic correlation time in the high collision frequency limit, the accuracy and efficiency of sampling are relatively insensitive to the value of the collision frequency in a broad range. After we compare the behaviors of the Andersen thermostat to those of Langevin dynamics, a heuristic schematic representation is proposed for understanding efficient stochastic thermostatting processes with molecular dynamics.

## I. Introduction

Molecular dynamics (MD) coupled with the stochastic process has offered a powerful tool for investigating structural and thermodynamic properties for such as the canonical ensemble, where the number of particles (*N*), the volume (*V*), and the temperature (*T*) are constant. (3*N* is the total number of degrees of freedom, which becomes one when a one-dimensional one-particle system is considered.) The stochastic process[1-3] serves as a type of thermostatting method to control the temperature of the system. Some prevailing stochastic processes include Langevin dynamics, the Andersen thermostat[4], *etc*. Because the time interval of MD is often finite in a computer simulation, the underlying numerical algorithm for MD with the stochastic process affects not only the accuracy but also the efficiency of the sampling (of the ensemble).

Langevin dynamics is a type of stochastic thermostat for MD for sampling constant temperature ensembles. Its equations of motion are described by the Langevin equation[5]

$$d\mathbf{x}_t = \mathbf{M}^{-1}\mathbf{p}_t dt \quad , \tag{1}$$

$$d\mathbf{p}_t = -\nabla_{\mathbf{x}_t} U(\mathbf{x}_t)dt - \boldsymbol{\gamma}\mathbf{p}_t dt + \boldsymbol{\sigma}\mathbf{M}^{1/2} d\mathbf{W}_t \quad , \tag{2}$$

where $\mathbf{M}$ is the diagonal "mass matrix" with elements $\{m_j\}$, and $\mathbf{p}_t$ and $\mathbf{x}_t$ are the momentum and coordinate vectors, respectively, $\mathbf{W}_t$ is a vector of 3*N*-dimensional independent Wiener processes, $\boldsymbol{\gamma}$ is often a diagonal friction matrix with positive elements, and $\boldsymbol{\sigma} = \sqrt{2/\beta}\boldsymbol{\gamma}^{1/2}$ . (The inverse temperature $\beta = 1/k_B T$ with $k_B$ as the Boltzmann constant.) Here and in the following a function *F* of time *t* ( $F(t)$ ) is also denoted as $F_t$ for abbreviation. The relation between the matrix $\boldsymbol{\sigma}$ and the friction matrix $\boldsymbol{\gamma}$ is defined by the fluctuation-dissipation theorem. This leads to the Boltzmann distribution $e^{-\beta H(\mathbf{x},\mathbf{p})}$ as the stationary state for Eqs. (1)-(2), where $H(\mathbf{x},\mathbf{p})$ is the (time-independent) Hamiltonian of the system

$$H = \mathbf{p}^T \mathbf{M}^{-1} \mathbf{p}/2 + U(\mathbf{x}) \quad . \tag{3}$$

In the literature various numerical algorithms were proposed to use Langevin dynamics to obtain the desired Boltzmann distribution. Leimkuhler and Matthews have recently compared a few numerical algorithms for Langevin dynamics in the high friction limit[6-8] for their performances in accuracy. In a more recent paper we have given a comprehensive study on various Langevin dynamics algorithms for the sampling efficiency as well as the accuracy in a broad range of the value of the friction coefficient[9]. In addition to the real dynamics case in the conventional theoretical framework of the Langevin equation, it is shown that virtual dynamics—a type of discrete evolution that may not correspond to a continuous, real dynamical counterpart of the equations of motion—is also able to yield the desired stationary distribution[9]. When the time interval $\Delta t$ is finite, the characteristic correlation time (of the potential or of the Hamiltonian) of such as the "middle" scheme of Langevin dynamics reaches the same plateau for both real and virtual dynamics in the high friction limit. Define the step number $\tau/\Delta t$ as the characteristic correlation time $\tau$ divided by the finite time interval $\Delta t$. The larger the time interval $\Delta t$ is, the smaller the step number of the value of the plateau is. This suggests that the result will be relatively insensitive to the value of the friction coefficient in a wide range.

The unified theoretical framework proposed in Ref. [10] indicates that these conclusions (for Langevin dynamics) can be applied to other stochastic processes for thermostats. In the work we use the Andersen thermostat[4, 10] as an example.

The paper is organized as follows: Section II briefly reviews the real and virtual dynamics cases of the Langevin thermostat[9]. In Section III we suggest that the Andersen thermostat also have the virtual dynamics case. In addition, we employ the phase space propagator approach[9] to derive the characteristic correlation time for the 1-dimensional harmonic systems for both the real and virtual dynamics cases of the Andersen thermostat. We focus the analysis on the "middle" scheme throughout the paper, since it is recommended for configurational sampling *via* molecular dynamics for the canonical ensemble[9, 10]. [If accurate phase space sampling is necessary, we may use the "middle" scheme to obtain the marginal distribution of the configuration (the coordinate) while sampling the Gaussian momentum distribution (the Maxwell distribution) by Monte Carlo[9].] While Section IV compares the analytic results on the characteristic correlation time of the Andersen thermostat to those of Langevin dynamics, Section V presents several typical numerical examples to verify the conclusions obtained from the analytic analysis (of the harmonic system). Conclusion remarks are given in Section VI. A schematic representation is presented in the appendices.

A uniform time interval (or step size) $\Delta t$ will be adopted throughout the paper.

## II. The "middle" scheme for Langevin dynamics

### A. Real and virtual dynamics cases

The Fokker-Planck equation[5] (or the forward Kolmogorov equation) for Langevin dynamics is

$$\frac{\partial}{\partial t}\rho = -\left(\mathbf{M}^{-1}\mathbf{p}\right)\cdot\nabla_{\mathbf{x}}\rho + \nabla_{\mathbf{x}}U(\mathbf{x})\cdot\nabla_{\mathbf{p}}\rho + \left(\boldsymbol{\gamma}\nabla_{\mathbf{p}}\right)\cdot\left(\mathbf{p}\rho\right) + \left(\frac{1}{2}\boldsymbol{\sigma}^{2}\mathbf{M}\nabla_{\mathbf{p}}\right)\cdot\nabla_{\mathbf{p}}\rho \quad . \tag{4}$$

We may recast Eq. (4) as $\partial\rho/\partial t = \mathcal{L}\rho$ with the relevant Kolmogorov operator

$$\mathcal{L}\rho = -\left(\mathbf{M}^{-1}\mathbf{p}\right)\cdot\nabla_{\mathbf{x}}\rho + \nabla_{\mathbf{x}}U(\mathbf{x})\cdot\nabla_{\mathbf{p}}\rho + \nabla_{\mathbf{p}}\cdot\left(\boldsymbol{\gamma}\mathbf{p}\rho\right) + \frac{1}{\beta}\nabla_{\mathbf{p}}\cdot\left(\boldsymbol{\gamma}\mathbf{M}\nabla_{\mathbf{p}}\rho\right) \quad . \tag{5}$$

It is straightforward to show that the Boltzmann distribution $e^{-\beta H(\mathbf{x},\mathbf{p})}$ is a stationary state[9] for

$$\partial\rho/\partial t = \mathcal{L}\rho = 0 \quad . \tag{6}$$

The strategy to design numerical algorithms is based on the repartition of Eqs. (1) and (2) as demonstrated in Section III of Ref. [9]. We will focus on the "middle" scheme as it is the most efficient one.

Consider the repartition of Eqs. (1) and (2)

$$\begin{bmatrix} d\mathbf{x}_t \\ d\mathbf{p}_t \end{bmatrix} = \underbrace{\begin{bmatrix} \mathbf{M}^{-1}\mathbf{p}_t \\ 0 \end{bmatrix} dt}_{\mathrm{x}} + \underbrace{\begin{bmatrix} 0 \\ -\nabla_{\mathbf{x}_t}U(\mathbf{x}_t) \end{bmatrix} dt}_{\mathrm{p}} + \underbrace{\begin{bmatrix} 0 \\ -\boldsymbol{\gamma}\mathbf{p}_t dt + \boldsymbol{\sigma}\mathbf{M}^{1/2} d\mathbf{W}_t \end{bmatrix}}_{\mathrm{T}} \quad . \tag{7}$$

Suppose that the system starts with $\left(\mathbf{x}(t),\mathbf{p}(t)\right)$ at time *t*. The two solutions corresponding to the first and second terms in the RHS of Eq. (7), respectively, are given by

$$\begin{bmatrix} \mathbf{x}(t+\Delta t) \\ \mathbf{p}(t+\Delta t) \end{bmatrix} = \begin{bmatrix} \mathbf{x}(t) + \mathbf{M}^{-1}\mathbf{p}(t)\Delta t \\ \mathbf{p}(t) \end{bmatrix} , \tag{8}$$

$$\begin{bmatrix} \mathbf{x}(t+\Delta t) \\ \mathbf{p}(t+\Delta t) \end{bmatrix} = \begin{bmatrix} \mathbf{x}(t) \\ \mathbf{p}(t) - \nabla U(\mathbf{x})\big|_{\mathbf{x}=\mathbf{x}(t)} \Delta t \end{bmatrix} \quad . \tag{9}$$

The third term is the Ornstein-Uhlenbeck (OU) process. The update relation of this term reads as

$$\begin{bmatrix} \mathbf{x}(t+\Delta t) \\ \mathbf{p}(t+\Delta t) \end{bmatrix} = \begin{bmatrix} \mathbf{x}(t) \\ e^{-\boldsymbol{\gamma}\Delta t}\mathbf{p}(t) + \sqrt{\frac{1}{\beta}}\mathbf{M}^{1/2}\left(\mathbf{1} - e^{-2\boldsymbol{\gamma}\Delta t}\right)^{1/2} \boldsymbol{\eta}(t,\Delta t) \end{bmatrix} \quad . \tag{10}$$

Here and in the following $\mathbf{1}$ denotes the unit matrix with a suitable dimension obvious in the context. $\boldsymbol{\eta}(t,\Delta t)$ for a fixed time *t* is a standard-Gaussian-random-number vector with zero mean

$$\left\langle \boldsymbol{\eta}(t,\Delta t)\right\rangle = 0 \tag{11}$$

and diagonal deviation matrix

$$\left\langle \boldsymbol{\eta}(t,\Delta t)\boldsymbol{\eta}^T(t,\Delta t)\right\rangle = \mathbf{1} \quad . \tag{12}$$

Note that $\boldsymbol{\eta}(t,\Delta t)$ is different for each time step.

The phase space propagators for the three terms [Eqs. (8)-(10)] are denoted as $e^{\mathcal{L}_{\mathbf{x}}\Delta t}$, $e^{\mathcal{L}_{\mathbf{p}}\Delta t}$, and $e^{\mathcal{L}_T\Delta t}$, respectively. That is, the relevant Kolmogorov operators are

$$\mathcal{L}_{\mathbf{x}}\rho = -\left(\mathbf{M}^{-1}\mathbf{p}\right)\cdot\nabla_{\mathbf{x}}\rho \quad , \tag{13}$$

$$\mathcal{L}_{\mathbf{p}}\rho = \nabla_{\mathbf{x}}U(\mathbf{x})\cdot\nabla_{\mathbf{p}}\rho \quad , \tag{14}$$

$$\mathcal{L}_T\rho = \nabla_{\mathbf{p}}\cdot\left(\boldsymbol{\gamma}\mathbf{p}\rho\right) + \frac{1}{\beta}\nabla_{\mathbf{p}}\cdot\left(\boldsymbol{\gamma}\mathbf{M}\nabla_{\mathbf{p}}\rho\right) \quad , \tag{15}$$

where $\rho$ is a density distribution in the phase space. Integration over time in $\partial\rho/\partial t = \mathcal{L}_T\rho$ from Eq. (15) for the OU process, one may verify that

$$\begin{aligned} e^{\mathcal{L}_T\Delta t}\rho(\mathbf{x},\mathbf{p}) = {} & \left(\frac{\beta}{2\pi}\right)^{N/2}\left|\mathbf{M}\left(\mathbf{1}-e^{-2\boldsymbol{\gamma}\Delta t}\right)\right|^{-1/2}\int d\mathbf{p}_0\,\rho(\mathbf{x},\mathbf{p}_0) \\ & \times\exp\left[-\frac{\beta}{2}\left(\mathbf{p}-e^{-\boldsymbol{\gamma}\Delta t}\mathbf{p}_0\right)^T\mathbf{M}^{-1}\left(\mathbf{1}-e^{-2\boldsymbol{\gamma}\Delta t}\right)^{-1}\left(\mathbf{p}-e^{-\boldsymbol{\gamma}\Delta t}\mathbf{p}_0\right)\right] \end{aligned} \quad . \tag{16}$$

It is straightforward to show that the OU process keeps the Maxwell momentum distribution unchanged, i.e.,

$$e^{\mathcal{L}_T\Delta t}\exp\left\{-\beta\left[\frac{1}{2}\mathbf{p}^T\mathbf{M}^{-1}\mathbf{p}\right]\right\} = \exp\left\{-\beta\left[\frac{1}{2}\mathbf{p}^T\mathbf{M}^{-1}\mathbf{p}\right]\right\} \quad . \tag{17}$$

As we have demonstrated in Section IV-A of Ref. [9], replacing $e^{-\boldsymbol{\gamma}\Delta t}$ by $-e^{-\boldsymbol{\gamma}\Delta t}$ and $e^{-n\boldsymbol{\gamma}\Delta t}$ by $\left(-e^{-\boldsymbol{\gamma}\Delta t}\right)^n$ for any integer $n$ in Eq. (10) do not change the Maxwell momentum distribution, which also satisfies Eq. (17). That is, Eq. (10) becomes

$$\begin{bmatrix}\mathbf{x}(t+\Delta t)\\ \mathbf{p}(t+\Delta t)\end{bmatrix} = \begin{bmatrix}\mathbf{x}(t)\\ -e^{-\boldsymbol{\gamma}\Delta t}\mathbf{p}(t) + \sqrt{\frac{1}{\beta}}\mathbf{M}^{1/2}\left(\mathbf{1}-e^{-2\boldsymbol{\gamma}\Delta t}\right)^{1/2}\boldsymbol{\eta}(t,\Delta t)\end{bmatrix} \quad , \tag{18}$$

which presents another solution to the OU process [the third term in Eq. (7)], albeit not a physical solution. This is the *virtual* dynamics case that we obtained in Ref. [9]. We use $e^{\mathcal{L}_T^{vir}\Delta t}$ to denote the phase space propagator for Eq. (18). Eq. (16) then becomes

$$
\begin{aligned}
e^{\mathcal{L}_T^{vir}\Delta t}\rho(\mathbf{x},\mathbf{p}) = \left(\frac{\beta}{2\pi}\right)^{N/2} \left|\mathbf{M}\left(\mathbf{1}-e^{-2\gamma\Delta t}\right)\right|^{-1/2} \int d\mathbf{p}_0 \rho(\mathbf{x},\mathbf{p}_0) \\
\times \exp\left[-\frac{\beta}{2}\left(\mathbf{p}+e^{-\gamma\Delta t}\mathbf{p}_0\right)^T \mathbf{M}^{-1}\left(\mathbf{1}-e^{-2\gamma\Delta t}\right)^{-1}\left(\mathbf{p}+e^{-\gamma\Delta t}\mathbf{p}_0\right)\right]
\end{aligned} \quad . \tag{19}
$$

Note that the phase space propagator $e^{\mathcal{L}_T^{vir}\Delta t}$ in the virtual dynamics case also keeps the Maxwell momentum distribution unchanged, i.e.,

$$
e^{\mathcal{L}_T^{vir}\Delta t} \exp\left\{-\beta\left[\frac{1}{2}\mathbf{p}^T\mathbf{M}^{-1}\mathbf{p}\right]\right\} = \exp\left\{-\beta\left[\frac{1}{2}\mathbf{p}^T\mathbf{M}^{-1}\mathbf{p}\right]\right\} \quad . \tag{20}
$$

**B. “Middle” scheme**

Different splitting orders for Eq. (7) lead to different algorithms. The “middle” scheme[9] employs

$$
e^{\mathcal{L}\Delta t} \approx e^{\mathcal{L}^{\mathrm{Middle}}\Delta t} = e^{\mathcal{L}_{\mathbf{p}}\Delta t/2} e^{\mathcal{L}_{\mathbf{x}}\Delta t/2} e^{\mathcal{L}_T^{(real/vir)}\Delta t} e^{\mathcal{L}_{\mathbf{x}}\Delta t/2} e^{\mathcal{L}_{\mathbf{p}}\Delta t/2} \quad , \tag{21}
$$

in which $e^{\mathcal{L}_T^{(real/vir)}\Delta t}$ represents either the real dynamics case $e^{\mathcal{L}_T\Delta t}$ or the virtual dynamics case $e^{\mathcal{L}_T^{vir}\Delta t}$. It is easy to verify that the virtual dynamics case in Eq. (21) leads to the desired Boltzmann distribution. While an efficient Langevin thermostat algorithm proposed by Leimkuhler and Matthews[6] is only the real dynamics case of the “middle” scheme, our recent work[9] includes both real and virtual dynamics cases of the “middle” scheme. Interestingly, although an efficient Langevin dynamics algorithm was originally proposed by Grønbech-Jensen and Farago without employing the Lie-Trotter splitting[11], we proved that it is equivalent to the “middle” scheme of Langevin dynamics. That is, the theoretical framework that Eq. (21) offers[9] naturally unites Grønbech-Jensen and Farago’s work[11] and the progress by using the Lie-Trotter splitting[6-8, 10, 12, 13].

In Section IV of Ref. [9] we have shown two approaches—the trajectory-based approach by directly solving the discrete equations of motion and the phase space propagator approach—to obtain the stationary state distribution of the "middle" scheme

$$\rho^{\mathrm{Middle}}(x,p)=\frac{1}{Z_N}\exp\left\{-\beta\left[\frac{1}{2}p^T\left(M-M\omega^2\frac{\Delta t^2}{4}\right)^{-1}p+\frac{1}{2}\left(x-x_{eq}\right)^T M\omega^2\left(x-x_{eq}\right)\right]\right\} \quad , \tag{22}$$

where $Z_N$ is the normalization constant, for the one-dimensional harmonic system

$$U(x)=\frac{1}{2}\left(x-x_{eq}\right)^T M\omega^2\left(x-x_{eq}\right) \quad , \tag{23}$$

where $x_{eq}$ is a constant. The stationary configurational distribution is exact in the harmonic limit. Note that both the real and virtual dynamics cases lead to the same stationary state distribution.

Our recent work[10] indicates that the phase space propagator $e^{\mathcal{L}_T^{(real/vir)}\Delta t}$ may represent other thermostat processes rather than the real/virtual dynamics cases of Langevin dynamics. We will use the Andersen thermostat for the demonstration.

## III. Andersen thermostat

### A. Real and virtual dynamics cases

#### *1) Real dynamics case*

The Andersen thermostat[4] is a type of thermostat that employs the stochastic coupling to impose the desired temperature in the MD simulation. In the Andersen thermostat, each particle of the system stochastically collides with a fictitious heat bath, and once the collision occurs, the momentum of this particle is chosen afresh from the Maxwell-Boltzmann momentum distribution. Times between collisions with the heat bath are selected from a Poisson distribution, i.e., the probability distribution is $P(t;\nu)=\nu e^{-\nu t}$, where the collision frequency $\nu$ specifies the coupling strength between the particle and the heat bath. Between stochastic collisions, the propagation of the MD trajectory is at constant energy according to the Hamilton equations of motion or the Newtonian laws of motion. The collision step in the algorithm is often described as "Randomly select a number of particles to undergo a collision with the heat bath. The probability that a particle is selected in the time interval $\Delta t$ is $\nu\Delta t$ (more accurately, $1-e^{-\nu\Delta t}$). If particle *j* is selected, its new momentum is reselected from a Maxwell momentum distribution at the desired temperature *T*, while all other particles are unaffected by this collision." [14]

Note that the explicit form for the collision step at a time interval $\Delta t$ can be expressed as

$$\mathbf{p}^{(j)} \leftarrow \sqrt{\frac{1}{\beta}}\mathbf{M}_j^{1/2}\boldsymbol{\theta}_j, \quad \text{if } \mu_j < \nu\Delta t \left(\text{or more precisely } \mu_j < 1-e^{-\nu\Delta t}\right) \quad \left(j=\overline{1,N}\right) \quad . \qquad (24)$$

Here $\mathbf{p}^{(j)}$ is the 3-dimensional momentum vector and $\mathbf{M}_j$ the $3\times3$ diagonal mass matrix for particle *j*. $\mu_j$ is a uniformly distributed random number in the range (0,1), which is different for each particle $\left(j=\overline{1,N}\right)$, and each time when Eq. (24) is invoked. $\boldsymbol{\theta}_j$ is a 3-dimensional vector. Its element $\theta_j^{(i)}(t)$ is an independent Gaussian-distributed random number with zero mean and unit variance, which is different for each of three degrees of freedom (i.e., *x*, *y*, or *z*) in the 3-dimensional space $(i=1,2,3)$, each particle $\left(j=\overline{1,N}\right)$, and each invocation of Eq. (24).

As we have shown in Ref.[10], the Kolmogorov operator for the Andersen thermostat satisfies

$$\mathcal{L}\rho = \nu\left[\rho_{\mathrm{MB}}(\mathbf{p})\int_{-\infty}^{\infty}\rho(\mathbf{x},\mathbf{p})\,d\mathbf{p} - \rho(\mathbf{x},\mathbf{p})\right] - \left(\mathbf{M}^{-1}\mathbf{p}\right)\cdot\nabla_{\mathbf{x}}\rho + \nabla_{\mathbf{x}}U(\mathbf{x})\cdot\nabla_{\mathbf{p}}\rho \quad . \qquad (25)$$

Here $\rho_{\mathrm{MB}}(\mathbf{p})$ is the Maxwell (or Maxwell-Boltzmann) momentum distribution

$$\rho_{\mathrm{MB}}(\mathbf{p}) = \left(\frac{\beta}{2\pi}\right)^{3N/2} |\mathbf{M}|^{-1/2} \exp\left[-\frac{\beta}{2}\mathbf{p}^T\mathbf{M}^{-1}\mathbf{p}\right] \quad . \tag{26}$$

It is straightforward to show that the Boltzmann distribution $e^{-\beta H(\mathbf{x},\mathbf{p})}$ is a stationary state solution to Eq. (6) with the full Kolmogorov operator given by Eq. (25). I.e., the Andersen thermostat is able to generate the canonical ensemble (if ergodicity is guaranteed), a well-known statement from Refs. [3, 4, 14].

Use $e^{\mathcal{L}_T\Delta t}$ to represent the phase space propagator for the thermostat step at a time interval $\Delta t$ . Propagation of the density distribution in the phase space $\rho \equiv \rho(\mathbf{x},\mathbf{p},t)$ for the collision process can be characterized by the forward Kolmogorov equation

$$\frac{\partial \rho}{\partial t} = \mathcal{L}_T\rho = \nu\left[\rho_{\mathrm{MB}}(\mathbf{p})\int_{-\infty}^{\infty}\rho(\mathbf{x},\mathbf{p},t)\,d\mathbf{p} - \rho(\mathbf{x},\mathbf{p},t)\right] \quad . \tag{27}$$

Integration over time in Eq. (27) leads to

$$e^{\mathcal{L}_T\Delta t}\rho = (1-e^{-\nu\Delta t})\,\rho_{\mathrm{MB}}(\mathbf{p})\int_{-\infty}^{\infty}\rho(\mathbf{x},\mathbf{p})\,d\mathbf{p} + e^{-\nu\Delta t}\rho(\mathbf{x},\mathbf{p}) \quad . \tag{28}$$

It is much more convenient to use Eq. (27) or Eq. (28) to present the analytic analysis for the Andersen thermostat. Note that when $\nu\Delta t$ is small, an approximation of Eq. (28) yields

$$e^{\mathcal{L}_T\Delta t}\rho = \nu\Delta t\,\rho_{\mathrm{MB}}(\mathbf{p})\int_{-\infty}^{\infty}\rho(\mathbf{x},\mathbf{p})\,d\mathbf{p} + (1-\nu\Delta t)\,\rho(\mathbf{x},\mathbf{p}) \quad , \tag{29}$$

which corresponds to the conventional description for the collision step in the Andersen thermostat[4, 14].

***2) Virtual dynamics case***

The Andersen thermostat also has the virtual dynamics case. Consider the collision step in the Andersen thermostat described as follows,

Randomly select a number of particles to undergo a collision with the heat bath. The probability that a particle is selected in the time interval $\Delta t$ is $1-e^{-\nu\Delta t}$. If particle *j* is selected, its new momentum is reselected from a Maxwell momentum distribution at the desired temperature *T*. Otherwise, the new momentum of particle *j* takes the negative of its original value.

Note that the explicit form for the collision step at a time interval $\Delta t$ may be expressed as

$$\left.\begin{array}{ll}\mathbf{p}^{(j)} \leftarrow \sqrt{\frac{1}{\beta}}\mathbf{M}_j^{1/2}\boldsymbol{\theta}_j, & \text{if } \mu_j < 1-e^{-\nu\Delta t} \\ \mathbf{p}^{(j)} \leftarrow -\mathbf{p}^{(j)}, & \text{otherwise}\end{array}\right\} \left(j=\overline{1,N}\right) \quad . \tag{30}$$

Here $\mathbf{p}^{(j)}$, $\mathbf{M}_j$, and $\boldsymbol{\theta}_j$ have been defined in Eq. (24).

Use $e^{\mathcal{L}_T^{vir}\Delta t}$ to represent the phase space propagator for the thermostat step at a time interval $\Delta t$ in the virtual dynamics case described above. Propagation of the density distribution in the phase space $\rho \equiv \rho(\mathbf{x},\mathbf{p})$ for the collision process can be characterized by

$$e^{\mathcal{L}_T^{vir}\Delta t}\rho = \left(1-e^{-\nu\Delta t}\right)\rho_{\mathrm{MB}}(\mathbf{p})\int_{-\infty}^{\infty}\rho(\mathbf{x},\mathbf{p})d\mathbf{p} + e^{-\nu\Delta t}\rho(\mathbf{x},-\mathbf{p}) \quad . \tag{31}$$

For comparison, the phase space propagator $e^{\mathcal{L}_T\Delta t}$ (for the thermostat step at a time interval $\Delta t$) in the real dynamics case[10] leads to Eq. (28). Both the real and virtual dynamics cases of the Andersen thermostat satisfy the relation

$$e^{\mathcal{L}_T^{(real/vir)}\Delta t}\exp\left\{-\beta\left[\frac{1}{2}\mathbf{p}^T\mathbf{M}^{-1}\mathbf{p}\right]\right\} = \exp\left\{-\beta\left[\frac{1}{2}\mathbf{p}^T\mathbf{M}^{-1}\mathbf{p}\right]\right\} \quad . \tag{32}$$

That is, the stationary distribution with a finite time interval $\Delta t$ recovers the correct Boltzmann distribution in the free particle limit. This is a criterion for a *good* thermostat.

It is trivial to employ the phase space propagator approach in Appendix A of Ref. [10] and in Section IV of Ref. [9] to prove that both the real and virtual dynamics cases of the Andersen thermostat share the same stationary distribution [Eq. (22)] for the harmonic system when the “middle” scheme is employed.

## B. Characteristic correlation time in the harmonic limit

The sampling efficiency can be measured by the characteristic correlation time[9]. For example, the potential energy autocorrelation function is defined as

$$C_{pot}(t)=\frac{\left\langle\left[U\left(x(t)\right)-\left\langle U\left(x\right)\right\rangle\right]\left[U\left(x(0)\right)-\left\langle U\left(x\right)\right\rangle\right]\right\rangle}{\left\langle\left[U\left(x\right)-\left\langle U\left(x\right)\right\rangle\right]^2\right\rangle} \quad . \tag{33}$$

The bracket $\langle\ \rangle$ of Eq. (33) denotes the phase space average of the Boltzmann distribution. The characteristic correlation time for the potential correlation function is then given by

$$\tau_{pot}=\int_0^\infty C_{pot}(t)dt \quad . \tag{34}$$

The smaller $\tau_{pot}$ is, the more efficiently the thermostatting method explores the potential energy surface and samples the configurational space.

When the time interval $\Delta t$ is finite, the potential energy autocorrelation function [Eq. (33)] is expressed as

$$C_{pot}(n\Delta t)=\frac{\langle U(n\Delta t)U(0)\rangle-\langle U\rangle^2}{\langle U^2\rangle-\langle U\rangle^2} \quad . \tag{35}$$

The bracket $\langle\ \rangle$ of Eq. (35) denotes the phase space average of the stationary density distribution when the finite time interval $\Delta t$ is used. Its characteristic correlation time then becomes

$$\tau_{pot}=\Delta t\sum_{n=0}^{\infty}C_{pot}(n\Delta t) \quad . \tag{36}$$

Or the step number of its characteristic correlation time is

$$\tau_{pot}\big/\Delta t=\sum_{n=0}^{\infty}C_{pot}(n\Delta t) \quad . \tag{37}$$

Similarly, we can define the Hamiltonian autocorrelation function $C_{Ham}(n\Delta t)$ and its characteristic correlation time (for the finite time interval $\Delta t$ ) $\tau_{Ham}$ .

***1) Infinitesimal time interval***

It is easy to follow Section V-A of Ref. [9] and Appendix B of Ref. [10] to derive $\tau_{pot}$ and $\tau_{Ham}$ for the one-dimensional harmonic system [Eq. (23)] when the time interval is infinitesimal. The characteristic correlation time of the potential energy for an infinitesimal time interval for the Andersen thermostat is

$$\tau_{pot}^{\text{Andersen}} = \frac{1}{2}\left(\frac{2}{\nu} + \frac{\nu}{\omega^2}\right) , \tag{38}$$

for which the optimal value of the collision frequency is

$$\nu_{pot}^{\text{opt}} = \sqrt{2}\omega , \tag{39}$$

which produces the minimum characteristic correlation time

$$\tau_{pot}^{\text{Andersen, min}} = \frac{\sqrt{2}}{\omega} . \tag{40}$$

Analogously, we may obtain the characteristic correlation time of the Hamiltonian for an infinitesimal time interval for the Andersen thermostat

$$\tau_{Ham}^{\text{Andersen}} = \frac{2}{\nu} + \frac{\nu}{4\omega^2} , \tag{41}$$

for which the optimal value of the collision frequency is

$$\nu_{Ham}^{\text{opt}} = 2\sqrt{2}\omega \tag{42}$$

which yields the minimum characteristic correlation time

$$\tau_{Ham}^{\text{Andersen, min}} = \frac{\sqrt{2}}{\omega} . \tag{43}$$

***2) Finite time interval***

In Section V-B-1 of Ref. [9] we have shown how to use the phase space propagator approach to derive the characteristic correlation time of the potential and that of the Hamiltonian for the "middle" scheme of Langevin dynamics. The same approach can be employed to obtain the results for the Andersen thermostat in the "middle" scheme.

*a) Real dynamics case in the middle scheme*

The relevant Kolmogorov operators in Eq. (21) for the Andersen thermostat for the one-dimensional harmonic system [Eq. (23)] become

$$\mathcal{L}_x \rho = -\frac{p}{M}\frac{\partial \rho}{\partial x} \quad , \tag{44}$$

$$\mathcal{L}_p \rho = M\omega^2 \left(x - x_{eq}\right)\frac{\partial \rho}{\partial p} \quad , \tag{45}$$

$$\mathcal{L}_T \rho = \nu\left[\rho_{\mathrm{MB}}\left(p\right)\int_{-\infty}^{\infty}\rho\left(x,p\right)dp - \rho\left(x,p\right)\right] \quad . \tag{46}$$

We define the conditional densities

$$\begin{aligned}
&\rho_{n,0}\left(x,p\right) \equiv \rho\left(x,p;n\Delta t\left|x_0,p_0;0\right.\right) = \left(e^{\mathcal{L}^{\mathrm{Middle}}\Delta t}\right)^n \delta\left(x-x_0\right)\delta\left(p-p_0\right) \\
&\rho_{n,1}\left(x,p\right) \equiv e^{\mathcal{L}_p \Delta t/2}\rho_{n,0}\left(x,p\right) \\
&\rho_{n,2}\left(x,p\right) \equiv e^{\mathcal{L}_x \Delta t/2}\rho_{n,1}\left(x,p\right) \\
&\rho_{n,3}\left(x,p\right) \equiv e^{\mathcal{L}_T \Delta t}\rho_{n,2}\left(x,p\right) \\
&\rho_{n,4}\left(x,p\right) \equiv e^{\mathcal{L}_x \Delta t/2}\rho_{n,3}\left(x,p\right)
\end{aligned} \quad , \tag{47}$$

which lead to

$$\rho_{n+1,0}\left(x,p\right) = e^{\mathcal{L}_p \Delta t/2}\rho_{n,4}\left(x,p\right) \quad . \tag{48}$$

Although the explicit expression of $\rho_{n,i}\left(x,p\right)$ $\left(i=\overline{0,4}\right)$ is difficult to obtain, we directly analyze the displacement squared autocorrelation function

$$\begin{aligned}
&\left\langle \left(x_0 - x_{eq}\right)^2 \left(x_n - x_{eq}\right)^2 \right\rangle_i \\
&= \int \rho_0\left(x_0,p_0\right)\rho_{n,i}\left(x,p\right)\left(x_0 - x_{eq}\right)^2\left(x - x_{eq}\right)^2 dx_0 dp_0 dx dp \quad \left(i=\overline{0,4}\right)
\end{aligned} \quad . \tag{49}$$

When the "middle" scheme is employed, the initial distribution $\rho_0\left(x_0,p_0\right)$ is the stationary distribution [Eq. (22)] for the one-dimensional harmonic system [Eq. (23)].

Following the strategy in Section V-B-1 of Ref. [9], we may show

$$\boldsymbol{\chi}_{n,1} = \mathbf{G}_1 \boldsymbol{\chi}_{n,0} \quad , \tag{50}$$

$$\boldsymbol{\chi}_{n,2} = \mathbf{G}_2 \boldsymbol{\chi}_{n,1} \quad , \tag{51}$$

$$\boldsymbol{\chi}_{n,3} = \mathbf{G}_3 \boldsymbol{\chi}_{n,2} + \mathbf{g} \quad , \tag{52}$$

$$\boldsymbol{\chi}_{n,4} = \mathbf{G}_2 \boldsymbol{\chi}_{n,3} \quad , \tag{53}$$

$$\boldsymbol{\chi}_{n+1,0} = \mathbf{G}_1 \boldsymbol{\chi}_{n,4} \quad , \tag{54}$$

where

$$\boldsymbol{\chi}_{n,i} = \left( \left\langle \left(x_0 - x_{eq}\right)^2 \left(x_n - x_{eq}\right)^2 \right\rangle_i , \left\langle \left(x_0 - x_{eq}\right)^2 \left(x_n - x_{eq}\right) p_n \right\rangle_i , \left\langle \left(x_0 - x_{eq}\right)^2 p_n^2 \right\rangle_i \right)^T \quad \left(i = \overline{0,4}\right) \quad , \tag{55}$$

$$\mathbf{G}_1 = \begin{pmatrix} 1 & 0 & 0 \\ -M\omega^2 \dfrac{\Delta t}{2} & 1 & 0 \\ M^2\omega^4 \dfrac{\Delta t^2}{4} & -M\omega^2 \Delta t & 1 \end{pmatrix} , \tag{56}$$

$$\mathbf{G}_2 = \begin{pmatrix} 1 & \dfrac{\Delta t}{M} & \dfrac{\Delta t^2}{4M^2} \\ 0 & 1 & \dfrac{\Delta t}{2M} \\ 0 & 0 & 1 \end{pmatrix} , \tag{57}$$

$$\mathbf{G}_3 = \begin{pmatrix} 1 & 0 & 0 \\ 0 & e^{-\nu\Delta t} & 0 \\ 0 & 0 & e^{-\nu\Delta t} \end{pmatrix} , \tag{58}$$

and

$$\mathbf{g} = \left( 0, 0, \frac{1 - e^{-\nu\Delta t}}{\beta^2 \omega^2} \right)^T \quad . \tag{59}$$

Substituting Eqs. (50)-(53) into Eq. (54), we obtain

$$\boldsymbol{\chi}_{n+1,0} = \mathbf{G}_1 \mathbf{G}_2 \mathbf{G}_3 \mathbf{G}_2 \mathbf{G}_1 \boldsymbol{\chi}_{n,0} + \mathbf{G}_1 \mathbf{G}_2 \mathbf{g} \quad . \tag{60}$$

A more compact form of Eq. (60) is

$$\boldsymbol{\chi}_{n+1,0} = \bar{\mathbf{G}}\boldsymbol{\chi}_{n,0} + \bar{\mathbf{g}} \tag{61}$$

with

$$\bar{\mathbf{G}} = \mathbf{G}_1\mathbf{G}_2\mathbf{G}_3\mathbf{G}_2\mathbf{G}_1 \tag{62}$$

and

$$\bar{\mathbf{g}} = \mathbf{G}_1\mathbf{G}_2\mathbf{g} \quad . \tag{63}$$

When $n$ goes to infinity, $\rho_{n,0}\left(x,p\right)$ approaches the stationary distribution, i.e.,

$$\rho_{n,0}\left(x,p\right) \to \frac{1}{Z_N}\exp\left\{-\beta\left[\frac{p^2}{2M}\left(1-\frac{\omega^2\Delta t^2}{4}\right)^{-1} + \frac{1}{2}M\omega^2\left(x-x_{eq}\right)^2\right]\right\} \quad . \tag{64}$$

Then it is straightforward to verify

$$\begin{aligned}
&\left(\left\langle\left(x-x_{eq}\right)^2\right\rangle_0^2, \left\langle\left(x-x_{eq}\right)^2\right\rangle_0 \left\langle\left(x-x_{eq}\right)p\right\rangle_0, \left\langle\left(x-x_{eq}\right)^2\right\rangle_0 \left\langle p^2\right\rangle_0\right)^T \\
&= \boldsymbol{\chi}_{\infty,0} \\
&= \left(\mathbf{1}-\bar{\mathbf{G}}\right)^{-1}\bar{\mathbf{g}} \\
&= \left(\frac{1}{\beta^2M^2\omega^4}, 0, \frac{1-\frac{\omega^2\Delta t^2}{4}}{\beta^2\omega^2}\right)^T
\end{aligned} \quad , \tag{65}$$

and

$$\begin{aligned}
&\boldsymbol{\chi}_{0,0} - \left(\mathbf{1}-\bar{\mathbf{G}}\right)^{-1}\bar{\mathbf{g}} \\
&= \boldsymbol{\chi}_{0,0} - \boldsymbol{\chi}_{\infty,0} \\
&= \left(\left\langle\left(x-x_{eq}\right)^4\right\rangle_0 - \left\langle\left(x-x_{eq}\right)^2\right\rangle_0^2, \ \left\langle\left(x-x_{eq}\right)^3 p\right\rangle_0 - \left\langle\left(x-x_{eq}\right)^2\right\rangle_0 \left\langle\left(x-x_{eq}\right)p\right\rangle_0, \right. \\
&\qquad \left. \left\langle\left(x-x_{eq}\right)^2 p^2\right\rangle_0 - \left\langle\left(x-x_{eq}\right)^2\right\rangle_0 \left\langle p^2\right\rangle_0\right)^T \\
&= \left(\frac{2}{\beta^2M^2\omega^4}, 0, 0\right)^T
\end{aligned} \tag{66}$$

Rearranging Eq. (61) leads to

$$\boldsymbol{\chi}_{n+1,0}-\left(\mathbf{1}-\bar{\mathbf{G}}\right)^{-1}\bar{\mathbf{g}}=\bar{\mathbf{G}}\left[\boldsymbol{\chi}_{n,0}-\left(\mathbf{1}-\bar{\mathbf{G}}\right)^{-1}\bar{\mathbf{g}}\right] \quad . \tag{67}$$

The recursion formula Eq. (67) leads to

$$\boldsymbol{\chi}_{n,0}-\left(\mathbf{1}-\bar{\mathbf{G}}\right)^{-1}\bar{\mathbf{g}}=\bar{\mathbf{G}}^{n}\left[\boldsymbol{\chi}_{0,0}-\left(\mathbf{1}-\bar{\mathbf{G}}\right)^{-1}\bar{\mathbf{g}}\right] \quad . \tag{68}$$

Summing over $n$ from 0 to infinity in both sides of Eq. (68) produces

$$\sum_{n=0}^{\infty}\left[\boldsymbol{\chi}_{n,0}-\left(\mathbf{1}-\bar{\mathbf{G}}\right)^{-1}\bar{\mathbf{g}}\right]=\left(\mathbf{1}-\bar{\mathbf{G}}\right)^{-1}\left[\boldsymbol{\chi}_{0,0}-\left(\mathbf{1}-\bar{\mathbf{G}}\right)^{-1}\bar{\mathbf{g}}\right] \quad . \tag{69}$$

Substituting Eqs. (65)-(66) into Eq. (69), we obtain

$$\begin{pmatrix} \sum_{n=0}^{\infty}\left(\left\langle\left(x_0-x_{eq}\right)^2\left(x_n-x_{eq}\right)^2\right\rangle_0-\left\langle\left(x-x_{eq}\right)^2\right\rangle_0^2\right) \\ \sum_{n=0}^{\infty}\left(\left\langle\left(x_0-x_{eq}\right)^2\left(x_n-x_{eq}\right)p_n\right\rangle_0-\left\langle\left(x-x_{eq}\right)^2\right\rangle_0\left\langle\left(x-x_{eq}\right)p\right\rangle_0\right) \\ \sum_{n=0}^{\infty}\left(\left\langle\left(x_0-x_{eq}\right)^2 p_n^2\right\rangle_0-\left\langle\left(x-x_{eq}\right)^2\right\rangle_0\left\langle p^2\right\rangle_0\right) \end{pmatrix}=\left(\mathbf{1}-\bar{\mathbf{G}}\right)^{-1}\begin{pmatrix} \frac{2}{\beta^2 M^2\omega^4} \\ 0 \\ 0 \end{pmatrix} \quad . \tag{70}$$

The characteristic correlation time of the potential for a finite time interval $\Delta t$ [Eqs. (35)-(36)] is

$$\tau_{pot}=\Delta t\sum_{n=0}^{\infty}\frac{\left\langle\left(x_0-x_{eq}\right)^2\left(x_n-x_{eq}\right)^2\right\rangle_0-\left\langle\left(x-x_{eq}\right)^2\right\rangle_0^2}{\left\langle\left(x-x_{eq}\right)^4\right\rangle_0-\left\langle\left(x-x_{eq}\right)^2\right\rangle_0^2} \quad . \tag{71}$$

Eqs. (65), (70), and (71) lead to

$$\tau_{pot}^{\text{Andersen-real}}=\left[\left(\mathbf{1}-\bar{\mathbf{G}}\right)^{-1}\right]_{11}\Delta t \quad . \tag{72}$$

Here $\left[\left(\mathbf{1}-\bar{\mathbf{G}}\right)^{-1}\right]_{11}$ represents the element in the 1st row and 1st column of the matrix $\left(\mathbf{1}-\bar{\mathbf{G}}\right)^{-1}$. Substituting Eqs. (56)-(58) and (62) into Eq. (72) yields the explicit form

$$\tau_{pot}^{\text{Andersen-real}}=\frac{\left(1-e^{-\nu\Delta t}\right)^2+\left(3+6e^{-\nu\Delta t}-e^{-2\nu\Delta t}\right)\left(\frac{\omega\Delta t}{2}\right)^2}{\omega^2\Delta t\left(1+e^{-\nu\Delta t}\right)\left(1-e^{-\nu\Delta t}\right)} \quad . \tag{73}$$

Interestingly, Eq. (73) indicates

$$\tau_{pot}^{\text{Andersen-real}}\xrightarrow{\nu\to 0^+}\infty \quad , \tag{74}$$

$$\tau_{pot}^{\text{Andersen-real}} \xrightarrow{\nu\to\infty} \frac{1+3\left(\frac{\omega\Delta t}{2}\right)^2}{\omega^2\Delta t} \quad . \tag{75}$$

Eq. (74) holds for both an infinitesimal time interval and a finite one. While in the limit $\nu \to \infty$ for an infinitesimal time interval the characteristic correlation time of the potential is infinite, that for a finite time interval is, however, a constant.

The optimal collision frequency for Eq. (73) is

$$\nu_{pot}^{\text{Andersen-real, opt}} = \frac{1}{\Delta t}\ln\left(\frac{4+\omega^2\Delta t^2+2\sqrt{2}\omega\Delta t\sqrt{4-\omega^2\Delta t^2}}{4-3\omega^2\Delta t^2}\right) \quad (\text{when } \omega\Delta t < \frac{2}{\sqrt{3}}) \tag{76}$$

such that the characteristic correlation time reaches the minimum value

$$\tau_{pot}^{\text{Andersen-real, min}} = \frac{\Delta t\left[-6\omega^4\Delta t^4+\omega^2\Delta t^2\left(16-3\sqrt{2}\omega\Delta t\sqrt{4-\omega^2\Delta t^2}\right)+4\left(8+5\sqrt{2}\omega\Delta t\sqrt{4-\omega^2\Delta t^2}\right)\right]}{-8\omega^4\Delta t^4+2\omega^2\Delta t^2\left(16+\sqrt{2}\omega\Delta t\sqrt{4-\omega^2\Delta t^2}\right)+8\sqrt{2}\omega\Delta t\sqrt{4-\omega^2\Delta t^2}} . \tag{77}$$

As $\Delta t \to 0$, Eq. (73), Eq. (76), and Eq. (77) approach Eq. (38), Eq. (39) and Eq. (40), respectively.

Similarly, the characteristic correlation time of the Hamiltonian for a finite time interval $\Delta t$ for the "middle" scheme may be shown as

$$\tau_{Ham}^{\text{Andersen-real}} = \frac{\left[\begin{array}{l}\left(1-e^{-\nu\Delta t}\right)^2+\left(9+22e^{-\nu\Delta t}+e^{-2\nu\Delta t}\right)\left(\frac{\omega\Delta t}{2}\right)^2 \\ -\left(9+22e^{-\nu\Delta t}+e^{-2\nu\Delta t}\right)\left(\frac{\omega\Delta t}{2}\right)^4+\left(3+6e^{-\nu\Delta t}-e^{-2\nu\Delta t}\right)\left(\frac{\omega\Delta t}{2}\right)^6\end{array}\right]}{\omega^2\Delta t\left(1+e^{-\nu\Delta t}\right)\left(1-e^{-\nu\Delta t}\right)\left[\left(1-\frac{\omega^2\Delta t^2}{4}\right)^2+1\right]} \quad . \tag{78}$$

Eq. (78) leads to

$$\tau_{Ham}^{\text{Andersen-real}} \xrightarrow{\nu\to 0^+} \infty \quad , \tag{79}$$

$$\tau_{Ham}^{\text{Andersen-real}} \overset{\nu\to\infty}{\to} \frac{1+9\left(\frac{\omega\Delta t}{2}\right)^2 - 9\left(\frac{\omega\Delta t}{2}\right)^4 + 3\left(\frac{\omega\Delta t}{2}\right)^6}{\omega^2\Delta t\left[\left(1-\frac{\omega^2\Delta t^2}{4}\right)^2+1\right]} \quad . \tag{80}$$

The characteristic correlation time of the Hamiltonian in the limit $\nu \to \infty$ for a finite time interval is also a constant.

The optimal collision frequency for Eq. (78) is

$$\nu_{Ham}^{\text{Andersen-real, opt}} = \frac{1}{\Delta t}\ln\left[\frac{64+80\omega^2\Delta t^2 - 20\omega^4\Delta t^4 + \omega^6\Delta t^6 + 2\sqrt{2}\omega\Delta t\left(8-\omega^2\Delta t^2\right)\left(4-\omega^2\Delta t^2\right)\sqrt{4-\omega^2\Delta t^2}}{64-176\omega^2\Delta t^2 + 44\omega^4\Delta t^4 - 3\omega^6\Delta t^6}\right] \tag{81}$$

$$\left(\text{when } \omega\Delta t < 0.634943\right)$$

such that the characteristic correlation time reaches the minimum value

$$\tau_{Ham}^{\text{Andersen-real, min}} = \frac{\varphi_1\Delta t}{\varphi_2} \tag{82}$$

with

$$\begin{aligned}\varphi_1 = {} & 6\omega^{16}\Delta t^{16} - 240\omega^{14}\Delta t^{14} + 4128\omega^{12}\Delta t^{12} - 39424\omega^{10}\Delta t^{10} + 223744\omega^8\Delta t^8 \\ & - 741376\omega^6\Delta t^6 + 1269760\omega^4\Delta t^4 - 655360\omega^2\Delta t^2 - 524288 \\ & + \left(3\omega^{10}\Delta t^{10} - 84\omega^8\Delta t^8 + 944\omega^6\Delta t^6 - 5184\omega^4\Delta t^4 + 14336\omega^2\Delta t^2 - 18432\right) \\ & \times\sqrt{2}\omega\Delta t\left(8-\omega^2\Delta t^2\right)\left(4-\omega^2\Delta t^2\right)\sqrt{4-\omega^2\Delta t^2}\end{aligned} \tag{83}$$

and

$$\begin{aligned}\varphi_2 = 32\left[\left(1-\frac{\omega^2\Delta t^2}{4}\right)^2+1\right]\Big[ & 4\omega^{12}\Delta t^{12} - 112\omega^{10}\Delta t^{10} + 1216\omega^8\Delta t^8 \\ & - 6400\omega^6\Delta t^6 + 16384\omega^4\Delta t^4 - 16384\omega^2\Delta t^2 \\ & + \left(-\omega^6\Delta t^6 + 20\omega^4\Delta t^4 - 80\omega^2\Delta t^2 - 64\right) \\ & \times\sqrt{2}\omega\Delta t\left(8-\omega^2\Delta t^2\right)\left(4-\omega^2\Delta t^2\right)\sqrt{4-\omega^2\Delta t^2}\Big]\end{aligned} \quad . \tag{84}$$

As $\Delta t \to 0$, Eq. (78), Eq. (81) and Eq. (82) approach Eq. (41), Eq. (42) and Eq. (43), respectively.

*b) Virtual dynamics case in the middle scheme*

Using the phase space propagator for the virtual dynamics version $e^{\mathcal{L}_T^{vir}\Delta t}$ in the "middle" scheme [Eq. (21)] leads to the "middle (vir)" scheme

$$e^{\mathcal{L}^{\text{Middle (vir)}}\Delta t} = e^{\mathcal{L}_p \Delta t/2} e^{\mathcal{L}_x \Delta t/2} e^{\mathcal{L}_T^{vir}\Delta t} e^{\mathcal{L}_x \Delta t/2} e^{\mathcal{L}_p \Delta t/2} \quad . \tag{85}$$

Note that Eq. (22) is also the stationary density distribution for the virtual dynamics case ["middle (vir)"] for the harmonic system.

Similar to Eq. (47), we have

$$\begin{aligned}
&\rho_{n,0}(x,p) \equiv \rho(x,p;n\Delta t|x_0,p_0;0) = \left(e^{\mathcal{L}^{\text{Middle (vir)}}\Delta t}\right)^n \delta(x-x_0)\delta(p-p_0) \\
&\rho_{n,1}(x,p) \equiv e^{\mathcal{L}_p \Delta t/2} \rho_{n,0}(x,p) \\
&\rho_{n,2}(x,p) \equiv e^{\mathcal{L}_x \Delta t/2} \rho_{n,1}(x,p) \\
&\rho_{n,3}(x,p) \equiv e^{\mathcal{L}_T^{vir}\Delta t} \rho_{n,2}(x,p) \\
&\rho_{n,4}(x,p) \equiv e^{\mathcal{L}_x \Delta t/2} \rho_{n,3}(x,p)
\end{aligned} \quad , \tag{86}$$

which lead to

$$\rho_{n+1,0}(x,p) = e^{\mathcal{L}_p \Delta t/2} \rho_{n,4}(x,p) \quad . \tag{87}$$

We define $\boldsymbol{\chi}_{i,n}$ in the same way as in the real dynamics case [Eq. (55)]. Analogously, we can verify

$$\begin{aligned}
\boldsymbol{\chi}_{n,1} &= \mathbf{G}_1 \boldsymbol{\chi}_{n,0} \\
\boldsymbol{\chi}_{n,2} &= \mathbf{G}_2 \boldsymbol{\chi}_{n,1} \\
\boldsymbol{\chi}_{n,3} &= \mathbf{G}_3' \boldsymbol{\chi}_{n,2} + \mathbf{g} \\
\boldsymbol{\chi}_{n,4} &= \mathbf{G}_2 \boldsymbol{\chi}_{n,3} \\
\boldsymbol{\chi}_{n+1,0} &= \mathbf{G}_1 \boldsymbol{\chi}_{n,4}
\end{aligned} \tag{88}$$

with $\mathbf{G}_1$, $\mathbf{G}_2$ and $\mathbf{g}$ defined in Eq. (56), Eq. (57), and Eq. (59), respectively, and

$$\mathbf{G}_3' = \begin{pmatrix} 1 & 0 & 0 \\ 0 & -e^{-\nu\Delta t} & 0 \\ 0 & 0 & e^{-\nu\Delta t} \end{pmatrix} \quad . \tag{89}$$

Eq. (88) leads to

$$\boldsymbol{\chi}_{n+1,0} = \bar{\mathbf{G}}'\boldsymbol{\chi}_{n,0} + \bar{\mathbf{g}} \tag{90}$$

with $\bar{\mathbf{G}}' = \mathbf{G}_1\mathbf{G}_2\mathbf{G}_3'\mathbf{G}_2\mathbf{G}_1$ and $\bar{\mathbf{g}} = \mathbf{G}_1\mathbf{G}_2\mathbf{g}$. Following the same procedure as in the real dynamics case, the characteristic correlation time of the potential for "middle (vir)" may be shown as

$$\tau_{pot}^{\text{Andersen-vir}} = \left[\left(\mathbf{1}-\bar{\mathbf{G}}'\right)^{-1}\right]_{11}\Delta t = \frac{\left(1+e^{-\nu\Delta t}\right)+\left(3-e^{-\nu\Delta t}\right)\left(\frac{\omega\Delta t}{2}\right)^2}{\omega^2\Delta t\left(1-e^{-\nu\Delta t}\right)} \quad . \tag{91}$$

Similarly, we obtain the characteristic correlation time of the Hamiltonian for "middle (vir)"

$$\tau_{Ham}^{\text{Andersen-vir}} = \frac{\left(1+e^{-\nu\Delta t}\right)+\left(9+e^{-\nu\Delta t}\right)\left(\frac{\omega\Delta t}{2}\right)^2-\left(9+e^{-\nu\Delta t}\right)\left(\frac{\omega\Delta t}{2}\right)^4+\left(3-e^{-\nu\Delta t}\right)\left(\frac{\omega\Delta t}{2}\right)^6}{\omega^2\Delta t\left(1-e^{-\nu\Delta t}\right)\left[\left(1-\frac{\omega^2\Delta t^2}{4}\right)^2+1\right]} \quad . \tag{92}$$

In the virtual dynamics case of the "middle" scheme, the characteristic correlation time of either the potential or the Hamiltonian monotonically decreases as the frequency $\nu$ increases.

It is easy to show

$$\tau_{pot}^{\text{Andersen-vir}} - \tau_{pot}^{\text{Andersen-real}} = \frac{4e^{-\nu\Delta t}\left(1-\frac{\omega^2\Delta t^2}{4}\right)}{\omega^2\Delta t\left(1+e^{-\nu\Delta t}\right)\left(1-e^{-\nu\Delta t}\right)} > 0 \quad , \tag{93}$$

$$\tau_{Ham}^{\text{Andersen-vir}} - \tau_{Ham}^{\text{Andersen-real}} = \frac{4e^{-\nu\Delta t}\left(1-\frac{\omega^2\Delta t^2}{4}\right)^3}{\omega^2\Delta t\left(1+e^{-\nu\Delta t}\right)\left(1-e^{-\nu\Delta t}\right)\left[\left(1-\frac{\omega^2\Delta t^2}{4}\right)^2+1\right]} > 0 \quad , \tag{94}$$

i.e., $\tau_{pot}^{\text{Andersen-vir}} > \tau_{pot}^{\text{Andersen-real}}$ and $\tau_{Ham}^{\text{Andersen-vir}} > \tau_{Ham}^{\text{Andersen-real}}$ when the collision frequency $\nu$ is finite. The characteristic correlation time for the virtual dynamics case is *always* larger than that for the real dynamics case. (Note $\omega\Delta t/2 < 1$ always holds in the stable region of the "middle" schem. See Appendix B.)

Interestingly, for a finite time interval $\Delta t$ we have

$$\tau_{pot}^{\text{Andersen-vir}} \overset{\nu\to\infty}{\to} \frac{1+3\left(\frac{\omega\Delta t}{2}\right)^2}{\omega^2\Delta t} \quad , \tag{95}$$

$$\tau_{Ham}^{\text{Andersen-vir}} \overset{\nu\to\infty}{\to} \frac{1+9\left(\frac{\omega\Delta t}{2}\right)^2-9\left(\frac{\omega\Delta t}{2}\right)^4+3\left(\frac{\omega\Delta t}{2}\right)^6}{\omega^2\Delta t\left\{\left[1-\left(\frac{\omega\Delta t}{2}\right)^2\right]^2+1\right\}} \quad . \tag{96}$$

That is, as $\nu \to \infty$ the characteristic correlation time for the virtual dynamics case approaches the same limit as that for the real dynamics case does.

In the limit $\nu \to 0^+$ the characteristic correlation time approach infinity, regardless of whether the real or virtual dynamics case of the Andersen thermostat is employed for a finite time interval $\Delta t$.

**IV. Comparison between Langevin dynamics and the Andersen thermostat in the "middle" scheme for the harmonic system**

In Appendix A the results on the characteristic correlation time for Langevin dynamics in the "middle" scheme[9] are briefly reviewed. It is suggested in Ref. [10] and in Appendix A that the collision frequency $\nu$ and $\sqrt{2}$ times of the friction coefficient $\gamma$ are two comparable parameters, i.e., Eq. (A7). Define a new parameter

$$\xi = \nu = \sqrt{2}\gamma \quad . \tag{97}$$

Regardless of the value of $\xi$, either of the Andersen thermostat and Langevin dynamics yields the stationary state distribution Eq. (22) for a finite time interval $\Delta t$ for the 1-dimensional harmonic system Eq. (23). That is, the accuracy is irrelevant to the thermostat parameter in the harmonic limit.

While Fig. 1 compares the Andersen thermostat to Langevin dynamics on the behavior of the characteristic correlation time of the potential as a function of the parameter $\xi$ defined in Eq. (97), Fig. 2 does so for the characteristic correlation time of the Hamiltonian. Figs. 1 and 2 demonstrate that the real dynamics case of Langevin dynamics always yields the smallest characteristic correlation time. The ascendant order for the value of the characteristic correlation time is

Langevin (real) < Andersen (real) < Andersen (virtual) < Langevin (virtual) (98)

when the time interval $\Delta t$ is relatively short. As $\Delta t$ increases, the order is changed. When $\Delta t$ is large enough, the ascendant order for the value of the characteristic correlation time becomes

Langevin (real) < Langevin (virtual) < Andersen (real) < Andersen (virtual). (99)

Interestingly, when the time interval $\Delta t$ is significantly large, the real dynamics case of the Andersen thermostat may even lead to a larger characteristic correlation time than the virtual dynamics case of Langevin dynamics!

Figs. 1 and 2 also show that Langevin dynamics and the Andersen thermostat share the same plateau of the characteristic correlation time in the high-friction/high-collision-frequency limit. This is consistent with the comparison of Eqs. (75), (80), (95)-(96) for the Andersen thermostat to the corresponding results for Langevin dynamics of Ref. [9] when the 1-dimensional harmonic system is studied. When $\nu$ approaches infinity in the Andersen thermostat, the probability for selecting a new momentum from the Maxwell momentum distribution at the desired temperature $T$ in the collision step approaches 1. This is effectively the same for the OU process of Langevin dynamics when $\gamma$ approaches infinity. Hence, the characteristic correlation time of the potential (or the Hamiltonian) in the limit $\nu \to \infty$ for the Andersen thermostat are expected to be the same with that in the limit $\gamma \to \infty$ for Langevin dynamics. The conclusion holds even for general anharmonic systems.

As demonstrated in Figs. 1 and 2, when $\Delta t$ is relatively large the characteristic correlation time is then relatively insensitive in a wide range of the thermostat parameter $\xi$. Such a range for Langevin dynamics is wider than that for the Andersen thermostat. The former also accesses relatively small values of the thermostat parameter $\xi$, as shown in panels (d) and (e) of Fig. 1 and in panels (c)-(e) of Fig. 2.

When the real dynamics case is employed, the optimal friction coefficient that yields the minimum characteristic correlation time is often a function of the time interval $\Delta t$. When the characteristic correlation time monotonically decays as the friction coefficient increases, the optimal friction coefficient becomes infinite. Panels (a) and (c) of Fig. 3 show the minimum characteristic correlation time as a function of the time interval $\Delta t$. Langevin dynamics and the Andersen thermostat are comparable on the sampling efficiency when the optimal thermostat parameters are used. Panels (b) and (d) of Fig. 3 depict the optimal friction coefficient as a function of the time interval $\Delta t$. For the Andersen thermostat, the characteristic correlation time of the potential as a function of the collision frequency has a minimum in the region $\omega \Delta t < \frac{2}{\sqrt{3}}$ [suggested by Eq. (76)], while that of the Hamiltonian has a minimum only when $\omega \Delta t < 0.634943$ [indicated by Eq. (81)]. In contrast, the characteristic correlation time of Langevin dynamics always has a minimum before the real dynamics breaks down. When $\Delta t$ is fixed, comparing to the Andersen thermostat, Langevin dynamics has a wider range of the thermostat parameter $\xi$ in which the sampling efficiency is insensitive. This is consistent with our discussion on Figs. 1 and 2.

Although Langevin dynamics and the Andersen thermostat share the same plateau in the high-friction/high-collision-frequency limit, they lead to different minimal characteristic correlation times. Fig. 4 demonstrates the step number of the minimal characteristic correlation time $\tau_{pot}^{\min}/\Delta t$ (or $\tau_{Ham}^{\min}/\Delta t$) or the step number of the value of plateau $\tau_{pot}^{\xi\to\infty}/\Delta t$ (or $\tau_{Ham}^{\xi\to\infty}/\Delta t$) as a function of the time interval $\Delta t$. The step number of the characteristic correlation time is more useful for describing the discrete evolution in the computer simulation. As depicted in Fig. 4, the step number of the value of the plateau monotonically decreases as the time interval $\Delta t$ increases, so is the step number of the minimal characteristic correlation time.

Analytic results for general anharmonic systems are often difficult to obtain if not possible. To verify the conclusions drawn from the analysis for the harmonic system we do the investigation with several numerical examples.

## V. Numerical examples

### A. Simulation details

We perform numerical simulations for the three typical anharmonic models of Ref. [9], for which the numerical performance of the Andersen thermostat is compared to that of Langevin dynamics.

The first model is the 1-dimensional quartic potential $U(x) = x^4/4$ (with the mass $m = 1$ and the inverse temperature $\beta = 1$). It contains no harmonic term and then presents a challenging model for testing the performance. We use the time intervals $\Delta t = 0.3$, 0.4, and 0.45 (unit: a.u.) for the accuracy of the result and $\Delta t = 0.1$, 0.3, and 0.4 (unit: a.u.) for the characteristic correlation time to test how their behaviors vary with the thermostat parameter.

In addition, two "real" molecular systems are investigated. The first example is a $H_2O$ molecule with the accurate potential energy surface (PES) developed by Partridge and Schwenke from extensive *ab initio* calculations and experimental data[15]. As the explicit form of the PES is available, that of the force may be expressed. This model is a good example for coupled intramolecular interactions. The MD simulations are performed for $T = 100$ K. While the time intervals $\Delta t = 1.9$, 2.2 and 2.4 (unit: fs) are used for testing the accuracy as a function of the thermostat parameter, $\Delta t = 0.24$, 1.2, 2.4 (unit: fs) are employed for examining the behavior of the characteristic correlation time as the thermostat parameter varies. The collision frequency $\nu$ ranges from $4.1 \times 10^{-3}$ fs$^{-1}$ to $4.2 \times 10^{4}$ fs$^{-1}$. After the system approaches equilibrium, 20 trajectories with each propagated up to 24 ns are used for estimating thermodynamic properties (the average potential energy and the thermal fluctuation of the potential are used as examples), the characteristic correlation time of the potential energy, and that of the Hamiltonian.

The second example is $(Ne)_{13}$, a Lennard-Jones (LJ) cluster. The parameters of the system are described in Ref. [16]. This model is a good example for coupled intermolecular interactions. The simulations are performed at $T = 14$ K. The time intervals $\Delta t = 20$, 50 (unit: fs) are used for computing thermodynamic properties and characteristic correlation times. The collision frequency ranges from $10^{-5}$ to $10^{3}$ fs$^{-1}$. After equilibrating the system, we employ 20 trajectories with each propagated up to 500 ns for estimating the average potential energy, the thermal fluctuation of the potential, the characteristic correlation time of the potential energy, and that of the Hamiltonian. We note that it is difficult to equilibrate this system with the virtual dynamics case of either the Andersen thermostat or Langevin dynamics when the thermostat parameter $\xi$ is smaller than $10^{-2}\ \mathrm{fs}^{-1}$.

For comparison, we also present the numerical results yielded by Langevin dynamics for these three systems, which are obtained from our earlier work[9].

**B. Results and discussions**

***1) Performance of real and virtual dynamics cases of the Andersen thermostat***

*a) Dependency of the numerical accuracy on the collision frequency*

We investigate two coordinate-dependent properties—the average potential energy $\langle U(\mathbf{x})\rangle$ and the thermal fluctuation of the potential $\sqrt{\langle U(\mathbf{x})^2\rangle - \langle U(\mathbf{x})\rangle^2}$, which indicate how the accuracy of the configurational sampling depends on the collision frequency.

As shown in Figs. 5-6 for the three typical systems, when the time interval $\Delta t$ is fixed the numerical results reach a plateau as long as the collision frequency is reasonably large. That is, the results are insensitive to the collision frequency in a broad region, irrespective of whether the real or virtual dynamics case is employed in the Andersen thermostat. The value of the plateau approaches the converged result as the time interval $\Delta t$ decreases.

*b) Dependency of the characteristic correlation time on the collision frequency*

We then consider the characteristic correlation time of the potential or that of the Hamiltonian, which represents the efficiency for sampling the configurational space or the phase space. As shown in Figs. 7-8, regardless of whether the real or virtual dynamics case is involved in the Andersen thermostat, the characteristic correlation time goes to infinity as the collision frequency approaches zero, while it gradually reaches a plateau as the collision frequency approaches infinity. Real and virtual dynamics share the same plateau in the high collision frequency limit when the same time interval is used. When the time interval $\Delta t$ is reasonably large without loss of much accuracy, the plateau value of the characteristic correlation time is considerably small, which indicates that it may be efficient and robust within a wide range of the collision frequency.

When the time interval is reasonably small, the characteristic correlation time often has a minimum, for which an optimal collision frequency exists. As the time interval increases, the minimum may disappear, i.e., the characteristic correlation time monotonically decays as the collision frequency increases. In contrast, there is no optimal collision frequency for the virtual dynamics case of the Andersen thermostat. The characteristic correlation time monotonically decays from infinity to the plateau as the collision frequency increases.

***2) Comparison of the Andersen thermostat to Langevin dynamics***

*a) Numerical accuracy*

Figs. 9-10 compare the Andersen thermostat to Langevin dynamics on the accuracy of the numerical results when the same time interval $\Delta t$ is used. In a broad range of the thermostat parameter, both the Andersen thermostat and Langevin dynamics lead to the same results (within statistical error bars). This is consistent with our earlier investigation in Ref. [10]. In terms of numerical accuracy, Langevin dynamics and the Andersen thermostat are comparable.

*b) Sampling efficiency*

Figs. 11-13 demonstrate that the ascendant order of the characteristic correlation time in Eq. (98) is also often valid in most cases of the three typical systems. The only exception is panel (d) of Fig. 12, where the characteristic correlation time of the Hamiltonian produced by the virtual dynamics case of the Andersen thermostat is larger than that yielded by the virtual

dynamics case of Langevin dynamics. In panel (d) of Fig. 12, the time interval $\Delta t$ is considerably large such that the ascendant order varies. This is consistent with our previous analysis on the change of the ascendant order in the harmonic system as shown in Figs. 1-2.

Figs. 11-13 show that the real dynamics case of Langevin dynamics always has a minimum characteristic correlation time as the thermostat parameter varies. In contrast, the real dynamics case of the Andersen thermostat produces a minimum value for the characteristic correlation time when the time interval $\Delta t$ is relatively small [e.g., panels (a)-(b) of Fig. 12], and does not do so when $\Delta t$ is large [e.g., panels (c)-(d) of Fig. 12].

In summary, in terms of sampling efficiency in most cases that we have investigated, the real dynamics case of Langevin dynamics is the most efficient while the virtual dynamics case of Langevin dynamics is the least.

## VI. Conclusion remarks

We have shown that there exists another type of discrete evolution [Eq. (30)] in the Andersen thermostat that leads to the desired stationary distribution. This virtual dynamics case of the Andersen thermostat is an analogy to the one of Langevin dynamics that we obtained in Ref. [9].

Because it is demonstrated that the "middle" scheme offers the most accurate and robust algorithm for any type of thermostat[10], in the paper we focus on the Andersen thermostat in the "middle" scheme. To investigate the sampling efficiency as well as the accuracy, we employ the phase space propagator approach introduced in our recent work[9, 10] to do the analytic analysis for the 1-dimensional harmonic system [Eq. (23)] when the time interval $\Delta t$ is finite. In addition, numerical simulations are performed for anharmonic models and "real" molecular systems. All the model tests suggest that both the accuracy and sampling efficiency are insensitive to the collision frequency in a broad region for the real dynamics case of the Andersen thermostat.

It is shown that the Andersen thermostat and Langevin dynamics (in the “middle” scheme) demonstrate similar behaviors. The two types of stochastic thermostatting processes lead to the same results (within statistical error bars) in a wide range of the thermostat parameter when the time interval $\Delta t$ is finite. While the characteristic correlation time (that describes the sampling efficiency) goes to infinity as the thermostat parameter approaches zero, it gradually reaches a plateau as the thermostat parameter approaches infinity. The characteristic correlation time of the virtual dynamics case always monotonically decays as the thermostat parameter increases. Both the Andersen thermostat and Langevin dynamics lead to the same plateau, regardless of whether real or virtual dynamics is employed. The step number of the value of the plateau ($\tau_{pot}^{\xi\to\infty} / \Delta t$ or $\tau_{Ham}^{\xi\to\infty} / \Delta t$) decreases as the time interval $\Delta t$ increases.

Significant differences exist between the Andersen thermostat and Langevin dynamics (in the “middle” scheme). While real dynamics of the Langevin equation always has a minimal characteristic correlation time, real dynamics of the Andersen thermostat does not have such a minimum when the time interval $\Delta t$ is considerably large. [See Appendices B-C and Tables 1-2 for more discussion when the 1-dimensional harmonic system is considered.] In most cases presented in the paper, virtual dynamics of the Andersen thermostat is more efficient than that of Langevin dynamics. In contrast, when the real dynamics case is employed, Langevin dynamics always performs better than the Andersen thermostat in terms of sampling efficiency.

In addition to the Andersen thermostat and Langevin dynamics, the unified theoretical framework proposed in Ref. [10] and our investigation in the present work suggest that virtual dynamics should in principle exist in other types of stochastic thermostatting processes. It is expected that the phase space propagator approach and the strategies that we employ in the paper will also be useful for understanding molecular dynamics with other types of stochastic processes. It will be interesting to combine real dynamics and virtual dynamics in the thermostat to develop more efficient thermostatting methods for MD and for path integral MD[10, 12].

**Acknowledgements**

This work was supported by the Ministry of Science and Technology of China (MOST) Grants No. 2016YFC0202803 and No. 2017YFA0204901, the National Natural Science Foundation of China (NSFC) Grants No. 21373018 and No. 21573007, by the Recruitment Program of Global Experts, by Specialized Research Fund for the Doctoral Program of Higher Education No. 20130001110009, and by Special Program for Applied Research on Super Computation of the NSFC-Guangdong Joint Fund (the second phase) under Grant No. U1501501. We acknowledge the Beijing and Tianjin supercomputer centers, the High-performance Computing Platform of Peking University, and the Computing Platform of the Center for Life Science of Peking University for providing computational resources.

## Appendix A. Characteristic correlation time for the 1-dimensional harmonic system for Langevin dynamics

Below we briefly review the results on the characteristic correlation time for the 1-dimensional harmonic system [Eq. (23)] for Langevin dynamics.

### 1. Infinitesimal time interval

It is shown in Appendix A of Ref. [12] and in Section V of Ref. [9] that for Langevin dynamics the characteristic correlation time of the potential energy for an infinitesimal time interval is

$$\tau_{pot}^{\text{Langevin}} = \frac{1}{2}\left(\frac{1}{\gamma} + \frac{\gamma}{\omega^2}\right) \quad , \tag{A1}$$

for which the optimal value of the friction coefficient is

$$\gamma_{pot}^{\text{opt}} = \omega \quad , \tag{A2}$$

which yields the minimum characteristic correlation time

$$\tau_{pot}^{\text{Langevin, min}} = \frac{1}{\omega} \quad . \tag{A3}$$

It is also shown that the characteristic correlation time of the Hamiltonian for an infinitesimal time interval

$$\tau_{Ham}^{\text{Langevin}} = \frac{1}{\gamma} + \frac{\gamma}{4\omega^2} \quad , \tag{A4}$$

for which the optimal value of the friction coefficient

$$\gamma_{Ham}^{\text{opt}} = 2\omega \quad , \tag{A5}$$

which produces the minimum characteristic correlation time

$$\tau_{Ham}^{\text{Langevin, min}} = \frac{1}{\omega} \quad . \tag{A6}$$

Comparing Eq. (A2) and Eq. (A5) (for Langevin dynamics) to Eq. (39) and Eq. (42) (for the Andersen thermostat), respectively, indicates that the relation between the friction coefficient (for Langevin dynamics) and the collision frequency (for the Andersen thermostat) is

$$\nu \cong \sqrt{2}\gamma \quad . \tag{A7}$$

## 2. Finite time interval

In Section V-B of Ref. [9], it is derived that the characteristic correlation time of the potential energy for the real dynamics case of Langevin dynamics is

$$\tau_{pot}^{\text{Langevin-real}} = \frac{\left(1-e^{-\gamma\Delta t}\right)^2 + \left(1+e^{-\gamma\Delta t}\right)\left(3-e^{-\gamma\Delta t}\right)\left(\frac{\omega\Delta t}{2}\right)^2}{\omega^2\Delta t\left(1+e^{-\gamma\Delta t}\right)\left(1-e^{-\gamma\Delta t}\right)} \quad . \tag{A8}$$

The optimal friction coefficient for Eq. (A8)

$$\gamma_{pot}^{\text{Langevin-real, opt}} = \frac{1}{\Delta t}\ln\left(\frac{2+\omega\Delta t}{2-\omega\Delta t}\right) \tag{A9}$$

leads to the minimum characteristic correlation time

$$\tau_{pot}^{\text{Langevin-real, min}} = \frac{2+\omega\Delta t}{2\omega} \quad . \tag{A10}$$

Taking the limit $\Delta t \to 0$ we can easily obtain

$$\tau_{pot}^{\text{Langevin-real}} \overset{\Delta t\to 0}{\to} \frac{1}{2}\left(\frac{1}{\gamma}+\frac{\gamma}{\omega^2}\right) \quad , \tag{A11}$$

$$\gamma_{pot}^{\text{Langevin-real, opt}} \overset{\Delta t\to 0}{\to} \omega \quad , \tag{A12}$$

$$\tau_{pot}^{\text{Langevin-real, min}} \overset{\Delta t\to 0}{\to} \frac{1}{\omega} \quad , \tag{A13}$$

which are the results on the characteristic correlation time of the potential for the infinitesimal time interval. Similarly, we obtain the characteristic correlation time of the Hamiltonian for the real dynamics case of Langevin dynamics[9],

$$\tau_{Ham}^{\text{Langevin-real}} = \frac{\left(1-e^{-\gamma\Delta t}\right)^2 + \left(3+e^{-\gamma\Delta t}\right)^2 \left(\frac{\omega\Delta t}{2}\right)^2 - \left(3+e^{-\gamma\Delta t}\right)^2 \left(\frac{\omega\Delta t}{2}\right)^4 + \left(3-e^{-\gamma\Delta t}\right)\left(1+e^{-\gamma\Delta t}\right)\left(\frac{\omega\Delta t}{2}\right)^6}{\omega^2\Delta t\left(1+e^{-\gamma\Delta t}\right)\left(1-e^{-\gamma\Delta t}\right)\left\{\left[1-\left(\frac{\omega\Delta t}{2}\right)^2\right]^2+1\right\}} ,\text{(A14)}$$

$$\gamma_{Ham}^{\text{Langevin-real, opt}} = \frac{1}{\Delta t}\ln\left\{\frac{1+5\left(\frac{\omega\Delta t}{2}\right)^2 - 5\left(\frac{\omega\Delta t}{2}\right)^4 + \left(\frac{\omega\Delta t}{2}\right)^6 + \omega\Delta t\left[2-\left(\frac{\omega\Delta t}{2}\right)^2\right]\sqrt{1+\left(\frac{\omega\Delta t}{2}\right)^2-\left(\frac{\omega\Delta t}{2}\right)^4}}{\left[1-\left(\frac{\omega\Delta t}{2}\right)^2\right]^3}\right\} ,\text{(A15)}$$

and

$$\tau_{Ham}^{\text{Langevin-real, min}} = -\frac{\phi_1\Delta t}{\phi_2} \quad , \tag{A16}$$

with

$$\begin{aligned}\phi_1 = {} & 6\omega^{14}\Delta t^{14} - 192\omega^{12}\Delta t^{12} + 2400\omega^{10}\Delta t^{10} - 13696\omega^8\Delta t^8 \\ & + 23552\omega^6\Delta t^6 + 88064\omega^4\Delta t^4 - 294912\omega^2\Delta t^2 - 131072 \\ & + \left(-\omega^{10}\Delta t^{10} + 36\omega^8\Delta t^8 - 432\omega^6\Delta t^6 + 2112\omega^4\Delta t^4 - 2048\omega^2\Delta t^2 - 10240\right) \\ & \times\sqrt{\omega^2\Delta t^2\left(8-\omega^2\Delta t^2\right)^2\left(-\omega^4\Delta t^4 + 4\omega^2\Delta t^2 + 16\right)}\end{aligned} \tag{A17}$$

and

$$\begin{aligned}\phi_2 = {} & 32\left\{\left[1-\left(\frac{\omega\Delta t}{2}\right)^2\right]^2+1\right\}\left[-2\omega^4\Delta t^4 + 8\omega^2\Delta t^2 + 32 + \sqrt{\omega^2\Delta t^2\left(8-\omega^2\Delta t^2\right)^2\left(-\omega^4\Delta t^4+4\omega^2\Delta t^2+16\right)}\right] \\ & \times\left[\omega^6\Delta t^6 - 16\omega^4\Delta t^4 + 64\omega^2\Delta t^2 + 2\sqrt{\omega^2\Delta t^2\left(8-\omega^2\Delta t^2\right)^2\left(-\omega^4\Delta t^4+4\omega^2\Delta t^2+16\right)}\right]\end{aligned} . \text{(A18)}$$

In the limit $\Delta t \to 0$ we may verify

$$\tau_{Ham}^{\text{Langevin-real}} \xrightarrow{\Delta t\to 0} \frac{1}{\gamma} + \frac{\gamma}{4\omega^2} \quad , \tag{A19}$$

$$\gamma_{Ham}^{\text{Langevin-real, opt}} \xrightarrow{\Delta t\to 0} 2\omega \quad , \tag{A20}$$

and

$$\tau_{Ham}^{\text{Langevin-real, min}} \xrightarrow{\Delta t\to 0} \frac{1}{\omega} \quad . \tag{A21}$$

Consider the high friction limit $\gamma \to \infty$. Eq. (A8) and Eq. (A14) lead to

$$\tau_{pot}^{\text{Langevin-real}} \overset{\gamma\to\infty}{\to} \frac{1+3\left(\frac{\omega\Delta t}{2}\right)^2}{\omega^2\Delta t} \tag{A22}$$

and

$$\tau_{Ham}^{\text{Langevin-real}} \overset{\gamma\to\infty}{\to} \frac{1+9\left(\frac{\omega\Delta t}{2}\right)^2-9\left(\frac{\omega\Delta t}{2}\right)^4+3\left(\frac{\omega\Delta t}{2}\right)^6}{\omega^2\Delta t\left\{\left[1-\left(\frac{\omega\Delta t}{2}\right)^2\right]^2+1\right\}} \quad . \tag{A23}$$

Comparing Eq. (A22) and Eq. (A23) to Eq. (75) and Eq. (80), respectively, shows that the real dynamics case of Langevin dynamics and that of the Andersen thermostat produce the same plateau of the characteristic correlation time in the high friction/collision-frequency limit.

**Appendix B. Stability analysis for the harmonic system**

Consider the multi-dimensional harmonic system

$$U(\mathbf{x}) = \frac{1}{2}\left(\mathbf{x}-\mathbf{x}_{eq}\right)^T \mathbf{A}\left(\mathbf{x}-\mathbf{x}_{eq}\right) \quad , \tag{B1}$$

where $\mathbf{x}_{eq}$ is a constant vector and $\mathbf{A}$ a symmetric and positive-definite constant Hessian matrix. We have $\mathbf{A} \leftarrow M\omega^2$ in the one-dimensional case [Eq. (23)]. As demonstrated in Section IV of Ref. [9] and in Appendix A of Ref. [10], the stationary state distribution for the "middle" scheme for either Langevin dynamics or the Andersen thermostat is a Gaussian distribution, i.e.,

$$\rho(\mathbf{x},\mathbf{p}) = \frac{1}{Z}\exp\left[-\frac{1}{2}\left(\mathbf{R}-\bar{\mathbf{R}}\right)^T\left(\mathbf{W}^{-1}\right)\left(\mathbf{R}-\bar{\mathbf{R}}\right)\right] \quad , \tag{B2}$$

where $\mathbf{R} = (\mathbf{x},\mathbf{p})^T$, $\bar{\mathbf{R}} = \left(\mathbf{x}_{eq},0\right)^T$, $Z$ is a normalization constant, and

$$\mathbf{W}^{-1} = \beta\begin{pmatrix} \mathbf{A} & \mathbf{0} \\ \mathbf{0} & \left(\mathbf{M}-\mathbf{A}\frac{\Delta t^2}{4}\right)^{-1} \end{pmatrix} \quad . \tag{B3}$$

The stability condition (for the "middle" scheme) is that the fluctuation correlation matrix $\mathbf{W}$ should be positive-definite. This implies that the Andersen thermostat and Langevin dynamics share the same stability condition in the harmonic limit.

For demonstration purpose, we show the stability condition for Langevin dynamics, for both the one-dimensional and multi-dimensional cases of the harmonic system.

**1. One-dimensional case**

In Section IV-B-2 of Ref.[9], we have shown the trajectory-based approach to derive the stationary state distribution for one-dimensional harmonic system [Eq. (23)] for the "middle" scheme of Langevin dynamics. The stability condition that the fluctuation correlation matrix $\mathbf{W}$ should be positive-definite, is equivalent to $|\varepsilon_{1,2}|<1$ with $\varepsilon_{1,2}$ the eigenvalues of the matrix $\tilde{\mathbf{M}}$, a $2\times 2$ matrix defined in Eqs. (73) and (76)-(79) of Ref.[9]. Here we use the variables $\tilde{\mathbf{M}}$, $\varepsilon_{1,2}$, $T$, and $D$ that are defined in Eqs. (73), (76)-(79) and (101)-(102) for the real dynamics case of the "middle" scheme of Langevin dynamics in Section IV-B of Ref.[9]. Note that it is trivial to define the corresponding variables for the virtual dynamics case.

***(1) Real dynamics case***

As discussed in Section IV-B of Ref.[9], for the real dynamics case we have

$$\begin{aligned} &\varepsilon_{1,2} = \frac{1}{2}\left(T \pm \sqrt{T^2 - 4D}\right) , \\ &D = e^{-\gamma\Delta t} , \\ &T = \left(1 + e^{-\gamma\Delta t}\right)\left(1 - \frac{\omega^2 \Delta t^2}{2}\right) . \end{aligned} \tag{B4}$$

There are two situations to consider:

(a) $T^2 - 4D < 0$

$\varepsilon_1$ and $\varepsilon_2$ are complex numbers with nonzero imaginary parts, and

$$\left|\varepsilon_{1,2}\right| = \left|\frac{1}{2}\left(T \pm \sqrt{T^2 - 4D}\right)\right| = \left|\frac{1}{2}\left(T \pm i\sqrt{4D - T^2}\right)\right| = \frac{1}{2}\sqrt{T^2 + \left(4D - T^2\right)} = \sqrt{D} = e^{-\frac{1}{2}\gamma\Delta t} < 1 \quad . \quad \text{(B5)}$$

The stability condition is always satisfied for this situation.

(b) $T^2 - 4D \geq 0$

$\varepsilon_1$ and $\varepsilon_2$ are real numbers. Because $0 < \varepsilon_1\varepsilon_2 = D < 1$, it is easy to verify

$$\left|\varepsilon_{1,2}\right| < 1 \Leftrightarrow \left(1 - \varepsilon_1^2\right)\left(1 - \varepsilon_2^2\right) > 0 \quad , \quad \text{(B6)}$$

and

$$\left(1 - \varepsilon_1^2\right)\left(1 - \varepsilon_2^2\right) = \left(1 + \varepsilon_1\varepsilon_2\right)^2 - \left(\varepsilon_1 + \varepsilon_2\right)^2 = \left(D + 1\right)^2 - T^2 = \left(D + 1 - T\right)\left(D + 1 + T\right) \quad . \quad \text{(B7)}$$

It is trivial to verify that $D + 1 - T > 0$ always holds. Eq. (B4) also leads to

$$D + 1 + T = \left(1 + e^{-\gamma\Delta t}\right)\left(2 - \frac{\omega^2 \Delta t^2}{2}\right) \quad . \quad \text{(B8)}$$

Eqs. (B6)-(B8) yield the stability condition $D + 1 + T > 0$, i.e.,

$$\omega\Delta t < 2 \quad . \quad \text{(B9)}$$

Although Eq. (B9) is always satisfied in part (a), it is not always the case in part (b). Eq. (B9) is then the stability condition for the real dynamics case.

***(2) Virtual dynamics case***

For the virtual dynamics case we have

$$\begin{aligned} \varepsilon_{1,2} &= \frac{1}{2}\left(T' \pm \sqrt{T'^2 - 4D'}\right) \ , \\ D' &= -e^{-\gamma\Delta t} \ , \\ T' &= \left(1 - e^{-\gamma\Delta t}\right)\left(1 - \frac{\omega^2 \Delta t^2}{2}\right) \ . \end{aligned} \quad \text{(B10)}$$

Note that $T'^2 - 4D' \geq 0$ is always satisfied. Following the same strategy in the real dynamics case, we can obtain the stability condition $D' + 1 + T' > 0$. It is not difficult to find

$$D' + 1 + T' = \left(1 - e^{-\gamma\Delta t}\right)\left(2 - \frac{\omega^2 \Delta t^2}{2}\right) \quad . \tag{B11}$$

We then verify that the stability condition of the virtual dynamics case is the same with that of the real dynamics case, i.e.,

$$\omega \Delta t < 2 \quad . \tag{B12}$$

**2. Multi-dimensional case**

The fluctuation correlation matrix $\mathbf{W}$ in Eq. (B3) should be positive-definite, which requires that the matrix $\mathbf{M} - \mathbf{A}\frac{\Delta t^2}{4}$ is positive-definite. The characteristic frequencies of the multi-dimensional harmonic system satisfy

$$\mathbf{M}^{-1/2}\mathbf{A}\mathbf{M}^{-1/2} = \mathbf{T}\left(\boldsymbol{\omega}^2\right)\mathbf{T}^T \quad . \tag{B13}$$

Here $\left(\boldsymbol{\omega}^2\right) = \begin{pmatrix} \omega_1^2 & & \\ & \omega_2^2 & \\ & & \ddots \end{pmatrix}$ with $\omega_i$ the characteristic frequency of each degree of freedom, and $\mathbf{T}$ is an orthogonal matrix. Then it is easy to show

$$\mathbf{M} - \mathbf{A}\frac{\Delta t^2}{4} = \mathbf{M}^{1/2}\mathbf{T}\left[\mathbf{1} - \frac{\Delta t^2}{4}\left(\boldsymbol{\omega}^2\right)\right]\mathbf{T}^T\mathbf{M}^{1/2} \quad . \tag{B14}$$

That is, the stability condition is that the matrix $\mathbf{1} - \frac{\Delta t^2}{4}\left(\boldsymbol{\omega}^2\right)$ is positive-definite, i.e., $\omega_i \Delta t < 2$, for each degree of freedom *i*.

## Appendix C. Schematic representation of the optimal friction coefficient or the optimal collision frequency for the one-dimensional harmonic system

Consider the one-dimensional harmonic system [Eq. (23)]. It is heuristic to have a schematic representation for Langevin dynamics or for the Andersen thermostat, in which the optimal friction coefficient or the optimal collision frequency is plotted as a function of the time interval $\Delta t$.

**1. Langevin dynamics**

Different domains have been proposed before for two Langevin dynamics algorithms[11, 17]. Interestingly, it is shown in Appendix B of Ref. [9] that Grønbech-Jensen and Farago's algorithm[11] is equivalent to the real and virtual dynamics cases of the "middle" scheme for Langevin dynamics. This suggests that we can define two characteristic variables

$$\begin{aligned} z_1 &= \omega\Delta t/2 \\ y_1 &= \gamma^{\mathrm{GF}}/2\omega \end{aligned} \quad , \tag{C1}$$

where the scaled friction coefficient

$$\gamma^{\mathrm{GF}} = \frac{2}{\Delta t}(1+a)^{-1}(1-a),\ a = \begin{cases} e^{-\gamma\Delta t}, & \text{for the real dynamics case} \\ -e^{-\gamma\Delta t}, & \text{for the virtual dynamics case} \end{cases} . \tag{C2}$$

The "middle" scheme of Langevin dynamics has four domains as shown in Fig. 14. As shown in Appendix B, the stability condition for the "middle" scheme is $\omega\Delta t/2 < 1$. The stable and unstable domains are then divided by the line $\omega\Delta t/2 = 1$.

1) Unstable region: $\omega\Delta t/2 > 1$ ($z_1 > 1$).

2) Stable region: $\omega\Delta t/2 < 1$ ($z_1 < 1$).

Eq. (C2) may be recast into

$$\gamma^{\mathrm{GF}}\Delta t/2 = (1+a)^{-1}(1-a)\begin{cases} <1, & \text{for the real dynamics case} \\ >1, & \text{for the virtual dynamics case} \end{cases} . \tag{C3}$$

The stable region then includes two parts separated by the line $\gamma^{\mathrm{GF}}\Delta t/2 = 1$:

a) Virtual dynamics region: $\gamma^{\mathrm{GF}}\Delta t/2 > 1$, i.e., $y_1 > \frac{1}{2z_1}$.

b) Real dynamics region: $\gamma^{\mathrm{GF}}\Delta t/2 < 1$, i.e., $y_1 < \frac{1}{2z_1}$.

The real dynamics region may be further separated into two domains. Consider the eigenvalues $\varepsilon_{1,2}$ for the real dynamics case as shown in Eq. (B4). We may recast $\varepsilon_{1,2}$ as

$$\varepsilon_{1,2} = e^{-\frac{1}{2}\gamma\Delta t \pm \lambda} \quad , \tag{C4}$$

where

$$\lambda = \ln\left( \frac{T + \sqrt{T^2 - 4e^{-\gamma\Delta t}}}{2e^{-\frac{1}{2}\gamma\Delta t}} \right) \tag{C5}$$

with $T$ defined in Eq. (B4). Here we define $\psi$ as

$$\lambda = i\psi \quad . \tag{C6}$$

When $T^2 - 4e^{-\gamma\Delta t} < 0$, i.e., $\varepsilon_{1,2}$ are complex numbers with nonzero imaginary parts as shown in part 1-(1) of Appendix B, $\psi$ is a real number in the range $(0, \pi)$. When $T^2 - 4e^{-\gamma\Delta t} > 0$, $\psi$ is an imaginary number with negative imaginary part. Consider the position-displacement autocorrelation function for a finite time interval $\Delta t$

$$C_x(n\Delta t) = \frac{\left\langle \left(x(0) - x_{eq}\right)\left(x(n\Delta t) - x_{eq}\right)\right\rangle - \left\langle x - x_{eq}\right\rangle^2}{\left\langle \left(x - x_{eq}\right)^2 \right\rangle - \left\langle x - x_{eq}\right\rangle^2} \quad . \tag{C7}$$

One may follow the phase space propagator approach (Section III-B-2) to derive

$$C_x(n\Delta t) = e^{-\frac{1}{2}\gamma n\Delta t}\left[ \frac{1 - e^{-\gamma\Delta t}}{2e^{-\frac{1}{2}\gamma\Delta t}\sinh\lambda}\sinh\left(n\lambda\right) + \cosh\left(n\lambda\right) \right] \quad . \tag{C8}$$

We denote $I(\omega')$ as the discrete Fourier transform of the position autocorrelation function, i.e.,

$$I(\omega') = \sum_{n=0}^{\infty} C_x(n\Delta t) e^{-i\omega' n\Delta t} \cdot \Delta t \quad . \tag{C9}$$

Substituting Eq. (C8) into Eq. (C9), we obtain

$$I(\omega')=\frac{\Delta t}{2}\left[\left(1+\frac{1-e^{-\gamma\Delta t}}{2e^{-\frac{1}{2}\gamma\Delta t}\sinh\lambda}\right)\frac{1}{1-e^{-\frac{1}{2}\gamma\Delta t+\lambda-i\omega'\Delta t}}+\left(1-\frac{1-e^{-\gamma\Delta t}}{2e^{-\frac{1}{2}\gamma\Delta t}\sinh\lambda}\right)\frac{1}{1-e^{-\frac{1}{2}\gamma\Delta t-\lambda-i\omega'\Delta t}}\right]. \quad \text{(C10)}$$

We focus on the real part of $I(\omega')$. Using the variable $\psi$, the real part of $I(\omega')$ can be recast into

$$\mathrm{Re}\left[I(\omega')\right]=\frac{\Delta t}{2}\left\{\frac{2\sin\psi\left[1-e^{-\frac{1}{2}\gamma\Delta t}\cos\left(\psi-\omega'\Delta t\right)\right]+\left(1-e^{-\gamma\Delta t}\right)\sin\left(\psi-\omega'\Delta t\right)}{2\sin\psi\left[1-2e^{-\frac{1}{2}\gamma\Delta t}\cos\left(\psi-\omega'\Delta t\right)+e^{-\gamma\Delta t}\right]}+\frac{2\sin\psi\left[1-e^{-\frac{1}{2}\gamma\Delta t}\cos\left(\psi+\omega'\Delta t\right)\right]+\left(1-e^{-\gamma\Delta t}\right)\sin\left(\psi+\omega'\Delta t\right)}{2\sin\psi\left[1-2e^{-\frac{1}{2}\gamma\Delta t}\cos\left(\psi+\omega'\Delta t\right)+e^{-\gamma\Delta t}\right]}\right\}, \quad \text{(C11)}$$

regardless of whether $\psi$ is a real number or not. Eq. (C11) shows that $\mathrm{Re}\left[I(\omega')\right]$ is a periodic function of $\omega'$. We may focus on the period ($-\pi/\Delta t$ , $\pi/\Delta t$ ) of $\omega'$. We can then define the two domains of the real dynamics case as:

i) Underdamped region: There are two peaks for $\mathrm{Re}\left[I(\omega')\right]$ in the period ($-\pi/\Delta t$ , $\pi/\Delta t$ );

ii) Overdamped region: There is only one peak for $\mathrm{Re}\left[I(\omega')\right]$ in the period ($-\pi/\Delta t$ , $\pi/\Delta t$ ).

We may verify from Eq. (C11) that, the dividing line between the underdamped and overdamped regions is described by

$$\cos\psi+\cosh\left(\frac{\gamma\Delta t}{2}\right)=2 \quad , \quad \text{(C12)}$$

where

$$\cos\psi=\frac{e^{i\psi}+e^{-i\psi}}{2}=\frac{\varepsilon_1+\varepsilon_2}{2e^{-\frac{1}{2}\gamma\Delta t}}=\frac{T}{2e^{-\frac{1}{2}\gamma\Delta t}}=\cosh\left(\frac{\gamma\Delta t}{2}\right)\left(1-\frac{\omega^2\Delta t^2}{2}\right) \quad . \quad \text{(C13)}$$

That is,

$$\cosh\left(\frac{\gamma\Delta t}{2}\right)\left(1-\frac{\omega^2\Delta t^2}{4}\right)=1 \quad . \tag{C14}$$

It is not difficult to show that $\cosh\left(\frac{\gamma\Delta t}{2}\right)\left(1-\frac{\omega^2\Delta t^2}{4}\right)<1$ holds for the underdamped region and that $\cosh\left(\frac{\gamma\Delta t}{2}\right)\left(1-\frac{\omega^2\Delta t^2}{4}\right)>1$ holds for the overdamped region. For the equation of the dividing line we have

$$\begin{aligned} y_1^2 &= \left[\frac{1}{\omega\Delta t}\left(1+e^{-\gamma\Delta t}\right)^{-1}\left(1-e^{-\gamma\Delta t}\right)\right]^2 \\ &= \frac{1}{\omega^2\Delta t^2}\frac{\sinh^2\left(\frac{\gamma\Delta t}{2}\right)}{\cosh^2\left(\frac{\gamma\Delta t}{2}\right)} \\ &= \frac{1}{\omega^2\Delta t^2}\left[1-\left(1-\frac{\omega^2\Delta t^2}{4}\right)^2\right] \\ &= \frac{1}{4}\left(2-z_1^2\right) \end{aligned} \quad , \tag{C15}$$

i.e.,

$$y_1 = \frac{1}{2}\sqrt{2-z_1^2} \quad . \tag{C16}$$

Then the two regions can be represented as:

i) Underdamped domain: $y_1 < \frac{1}{2}\sqrt{2-z_1^2}$ .

ii) Overdamped domain: $y_1 > \frac{1}{2}\sqrt{2-z_1^2}$ .

(Apparently the definition of the underdamped or overdamped domain is not unique. One may also choose another type of auto-correlation function to define the domains in the real dynamics region.)

While the blue curve in panel (a) of Fig. 14 represents the equation that the two characteristic variables $z_1$ and $y_1$ satisfy when the friction coefficient produces the optimal value for the characteristic correlation time of the potential energy, the red curve (in the same panel) does so for the optimal characteristic correlation time of the Hamiltonian.

Consider the blue curve. When $\gamma$ takes the optimal value for the minimal characteristic correlation time of the potential energy [Eq. (A9)], Eq. (C2) becomes

$$\gamma^{\mathrm{GF}} = \omega \quad . \tag{C17}$$

That is,

$$y_1 = \gamma^{\mathrm{GF}} / 2\omega = \frac{1}{2} \quad , \tag{C18}$$

which is the equation that $z_1$ and $y_1$ satisfy for the blue curve. Similarly, we may obtain the equation for the red curve. When the optimal friction coefficient for the minimal characteristic correlation time of the Hamiltonian Eq. (A15) is used, Eq. (C2) becomes

$$\gamma^{\mathrm{GF}} = \frac{2}{\Delta t} \frac{8\left(\frac{\omega\Delta t}{2}\right)^2 - 8\left(\frac{\omega\Delta t}{2}\right)^4 + 2\left(\frac{\omega\Delta t}{2}\right)^6 + \omega\Delta t\left[2 - \left(\frac{\omega\Delta t}{2}\right)^2\right]\sqrt{1 + \left(\frac{\omega\Delta t}{2}\right)^2 - \left(\frac{\omega\Delta t}{2}\right)^4}}{2 + 2\left(\frac{\omega\Delta t}{2}\right)^2 - 2\left(\frac{\omega\Delta t}{2}\right)^4 + \omega\Delta t\left[2 - \left(\frac{\omega\Delta t}{2}\right)^2\right]\sqrt{1 + \left(\frac{\omega\Delta t}{2}\right)^2 - \left(\frac{\omega\Delta t}{2}\right)^4}} . \tag{C19}$$

Using the two variables of Eq. (C1), we recast Eq. (C19) into

$$y_1 = \frac{\left(2 - z_1^2\right)\left[2z_1 - z_1^3 + \sqrt{1 + z_1^2 - z_1^4}\right]}{2\left[1 + z_1^2 - z_1^4 + z_1\left(2 - z_1^2\right)\sqrt{1 + z_1^2 - z_1^4}\right]} \quad , \tag{C20}$$

which is the equation that $z_1$ and $y_1$ satisfy for the red curve.

### 2. The Andersen thermostat

Similar to Eq. (C1), we define two characteristic variables for the Andersen thermostat

$$\begin{aligned} z_2 &= \omega\Delta t / 2 \\ y_2 &= \nu^{\mathrm{A}} / 2\omega \end{aligned} \tag{C21}$$

where the scaled collision frequency

$$\nu^{\mathrm{A}} = \frac{2}{\Delta t}\left(1 + a_{\mathrm{A}}\right)^{-1}\left(1 - a_{\mathrm{A}}\right),\ a_{\mathrm{A}} = \begin{cases} e^{-\nu\Delta t}, & \text{for the real dynamics case} \\ -e^{-\nu\Delta t}, & \text{for the virtual dynamics case} \end{cases} . \tag{C22}$$

The “middle” scheme has four regions as shown in Fig. 14.

1) Unstable region: $\omega\Delta t/2 > 1$, i.e., $z_2 > 1$.

2) Stable region: $\omega\Delta t/2 < 1$, i.e., $z_2 < 1$.

The stable region may be divided into two parts by the line $\nu^{\mathrm{A}}\Delta t/2 = 1$:

a) Virtual dynamics region: $\nu^{\mathrm{A}}\Delta t/2 > 1$, i.e., $y_2 > \frac{1}{2z_2}$.

b) Real dynamics region: $\nu^{\mathrm{A}}\Delta t/2 < 1$, i.e., $y_2 < \frac{1}{2z_2}$.

Similar to part 1 of Appendix C (where Eq. (C7) is used for defining the domains), it is not difficult to show that the real dynamics region (of the Andersen thermostat) is also separated into two domains by the line $y_2 = \frac{1}{2}\sqrt{2 - z_2^2}$:

i) Underdamped domain: $y_2 < \frac{1}{2}\sqrt{2 - z_2^2}$.

ii) Overdamped domain: $y_2 > \frac{1}{2}\sqrt{2 - z_2^2}$.

The blue curve in panel (b) of Fig. 14 represents the equation that the two characteristic variables $z_2$ and $y_2$ satisfy when the collision frequency approaches the optimal value for the minimal characteristic correlation time of the potential energy. In the region $0 < \omega\Delta t < 2/\sqrt{3}$ (i.e., $0 < z_2 < 1/\sqrt{3}$), the optimal collision frequency for the minimal characteristic correlation time of the potential energy is finite [Eq. (76)]. Substituting Eq. (76) into Eq. (C22) leads to

$$\nu^{\mathrm{A}} = \frac{1}{\Delta t}\frac{4\omega^2\Delta t^2 + 2\sqrt{2}\omega\Delta t\sqrt{4 - \omega^2\Delta t^2}}{4 - \omega^2\Delta t^2 + \sqrt{2}\omega\Delta t\sqrt{4 - \omega^2\Delta t^2}} \quad . \tag{C23}$$

By using the two variables defined in Eq. (C21), we recast Eq. (C23) into

$$y_2 = \frac{2z_2 + \sqrt{2}\sqrt{1 - z_2^2}}{2 - 2z_2^2 + 2\sqrt{2}z_2\sqrt{1 - z_2^2}} \quad . \tag{C24}$$

In the region $2/\sqrt{3} \le \omega\Delta t < 2$ (i.e., $1/\sqrt{3} \le z_2 < 1$), the optimal collision frequency for the minimal characteristic correlation time of the potential energy is infinite. This leads to

$$y_2 = \nu^{\mathrm{A}} / 2\omega = \frac{2}{\Delta t} \Big/ 2\omega = \frac{1}{2z_2} \quad . \tag{C25}$$

That is, the equation that $z_2$ and $y_2$ satisfy for the blue curve is

$$y_2 = \begin{cases} \dfrac{2z_2 + \sqrt{2}\sqrt{1-z_2^2}}{2 - 2z_2^2 + 2\sqrt{2} z_2 \sqrt{1-z_2^2}}, \text{ (for } 0 < z_2 < 1/\sqrt{3}) \\ \dfrac{1}{2z_2}, \text{ (for } 1/\sqrt{3} \le z_2 < 1) \end{cases} . \tag{C26}$$

We then consider the red curve in panel (b) of Fig. 14 that depicts the equation that $z_2$ and $y_2$ satisfy when the collision frequency takes the optimal value for the minimal characteristic correlation time of the Hamiltonian [Eq. (81)]. Eq. (C22) then becomes

$$\nu^{\mathrm{A}} = \frac{2}{\Delta t} \frac{256\omega^2\Delta t^2 - 64\omega^4\Delta t^4 + 4\omega^6\Delta t^6 + 2\sqrt{2}\omega\Delta t\left(8-\omega^2\Delta t^2\right)\left(4-\omega^2\Delta t^2\right)\sqrt{4-\omega^2\Delta t^2}}{128 - 96\omega^2\Delta t^2 + 24\omega^4\Delta t^4 - 2\omega^6\Delta t^6 + 2\sqrt{2}\omega\Delta t\left(8-\omega^2\Delta t^2\right)\left(4-\omega^2\Delta t^2\right)\sqrt{4-\omega^2\Delta t^2}} \tag{C27}$$

in the region $0 < \omega\Delta t < 0.634943$ (i.e., $0 < z_2 < 0.3174715$). Substituting Eq. (C27) into Eq. (C21) yields

$$y_2 = \frac{\left(2-z_2^2\right)\left[4z_2 - 2z_2^3 + \sqrt{2}\left(1-z_2^2\right)^{3/2}\right]}{2\left[1 - 3z_2^2 + 3z_2^4 - z_2^6 + \sqrt{2} z_2 \left(1-z_2^2\right)^{3/2}\left(2-z_2^2\right)\right]} \quad . \tag{C28}$$

When $0.634943 < \omega\Delta t < 2$ (i.e., $0.3174715 < z_2 < 1$), the optimal collision frequency for the minimal characteristic correlation time of the Hamiltonian is infinite. The equation for the two variables $z_2$ and $y_2$ in the region is the same as Eq. (C25). We then obtain the equation for the red curve as

$$y_2 = \begin{cases} \dfrac{\left(2-z_2^2\right)\left[4z_2 - 2z_2^3 + \sqrt{2}\left(1-z_2^2\right)^{3/2}\right]}{2\left[1-3z_2^2+3z_2^4-z_2^6+\sqrt{2}z_2\left(1-z_2^2\right)^{3/2}\left(2-z_2^2\right)\right]}, \text{ (for } 0 < z_2 < 0.3174715) \\ \dfrac{1}{2z_2}, \text{ (for } 0.3174715 < z_2 < 1) \end{cases} . \quad \text{(C29)}$$

Table 2 presents the equations of the curves for Langevin dynamics and for the Andersen thermostat.

**Tables and Figures**

**Table 1.** The range of $\omega\Delta t$ when the optimal value of friction coefficient or collision frequency is finite for the one-dimensional harmonic system [Eq. (23)].

| Thermostat | | Range of $\omega\Delta t$ |
|---|---|---|
| Langevin | $\tau_{pot}$ | (0, 2) |
| | $\tau_{Ham}$ | (0, 2) |
| Andersen | $\tau_{pot}$ | (0, $2/\sqrt{3}$ ) |
| | $\tau_{Ham}$ | (0, 0.634943) |

**Table 2.** The equations of the curves for the Langevin dynamics and those for the Andersen thermostat in Fig. 14.

| Thermostat | | $y \sim z$ |
|---|---|---|
| Langevin | potential | $y_1 = \dfrac{1}{2}$ |
| | Hamiltonian | $y_1 = \dfrac{\left(2 - z_1^2\right)\left[2z_1 - z_1^3 + \sqrt{1 + z_1^2 - z_1^4}\right]}{2\left[1 + z_1^2 - z_1^4 + z_1\left(2 - z_1^2\right)\sqrt{1 + z_1^2 - z_1^4}\right]}$ |
| Andersen | potential | $y_2 = \begin{cases} \dfrac{2z_2 + \sqrt{2}\sqrt{1 - z_2^2}}{2 - 2z_2^2 + 2\sqrt{2}z_2\sqrt{1 - z_2^2}}, \text{ (for } 0 < z_2 < \dfrac{1}{\sqrt{3}}) \\ \dfrac{1}{2z_2}, \text{ (for } \dfrac{1}{\sqrt{3}} \le z_2 < 1) \end{cases}$ |
| | Hamiltonian | $y_2 = \begin{cases} \dfrac{\left(2 - z_2^2\right)\left[4z_2 - 2z_2^3 + \sqrt{2}\left(1 - z_2^2\right)^{3/2}\right]}{2\left[1 - 3z_2^2 + 3z_2^4 - z_2^6 + \sqrt{2}z_2\left(1 - z_2^2\right)^{3/2}\left(2 - z_2^2\right)\right]}, \\ \text{(for } 0 < z_2 < 0.3174715) \\ \dfrac{1}{2z_2}, \text{ (for } 0.3174715 < z_2 < 1) \end{cases}$ |

**Fig. 1**. (Color online) Analytic results for the characteristic correlation time of the potential energy for the harmonic system [Eq. (23)]. The curves depict the equations that $\tau_{pot}\omega$ and $\xi/\omega$ satisfy for different parameters $\omega\Delta t$. Here $\xi$ stands for $\sqrt{2}\gamma$ for Langevin dynamics and $\nu$ for the Andersen thermostat. (a) For $\omega\Delta t \to 0^+$. (b) For $\omega\Delta t = 0.2$. (c) For $\omega\Delta t = 0.634943$. (d) For $\omega\Delta t = 2/\sqrt{3}$. (e) For $\omega\Delta t = 1.9$. "Langevin-real" and "Andersen-real" represent the analytic result for the real dynamics case of Langevin dynamics and that of the Andersen thermostat, respectively; "Langevin-vir" and "Andersen-vir" stand for the analytic result for the virtual dynamics case of Langevin dynamics and that of the Andersen thermostat, respectively.

**Fig. 2**. (Color online) Same as Fig. 1, but for the characteristic correlation time of the Hamiltonian for the harmonic system [Eq. (23)].

**Fig. 3**. (Color online) Analytic results for both Langevin dynamics and the Andersen thermostat for the harmonic system [Eq. (23)]. (a) $\tau_{pot}^{\min}\omega$ as a function of $\omega\Delta t$, where $\tau_{pot}^{\min}$ is the minimum value of the characteristic correlation time of the potential. (b) $\xi_{pot}^{opt}/\omega$ as a function of $\omega\Delta t$, where $\xi_{pot}^{opt}$ stands for $\sqrt{2}\gamma_{pot}^{opt}$ for Langevin dynamics and $\nu_{pot}^{opt}$ for the Andersen thermostat. Here $\gamma_{pot}^{opt}$ is the optimal friction coefficient and $\nu_{pot}^{opt}$ is the optimal collision frequency for the characteristic correlation time of the potential. Panels (c) and (d) are the same as panels (a) and (b), respectively, but for the characteristic correlation time of the Hamiltonian.

**Fig. 4**. (Color online) Analytic results for both Langevin dynamics and the Andersen thermostat for the harmonic system [Eq. (23)]. (a) $\tau_{pot}^{\min}/\Delta t$ as a function of $\omega\Delta t$, where $\tau_{pot}^{\min}/\Delta t$ is the step number of the minimal characteristic correlation time of the potential. (b) The same

as panel (a), but for the step number of the minimal characteristic correlation time of the Hamiltonian. (c) $\tau^{\xi\to\infty}/\Delta t$ as a function of $\omega\Delta t$, where $\tau^{\xi\to\infty}$ is the plateau value of the characteristic correlation time of the potential or that of the Hamiltonian [Here $\xi$ stands for $\sqrt{2}\gamma$ for Langevin dynamics and $\nu$ for the Andersen thermostat.], and $\tau^{\xi\to\infty}/\Delta t$ represents the step number of the value of the plateau.

**Fig. 5**. (Color online) Results for the average potential energy, the average Hamiltonian, the thermal fluctuation of the potential, and that of the Hamiltonian of the 1-dimensional quartic potential $U(x)=x^4/4$ at $\beta=1$ using different collision frequencies $\nu$ for the Andersen thermostat. Panels (a) and (b) present the average potential energy and Hamiltonian, respectively. Panels (c) and (d) show the thermal fluctuation of the potential energy and that of the Hamiltonian, respectively. Three time intervals $\Delta t = 0.3$, 0.4, and 0.45 are used. All the parameters are in atomic units (a.u.). "real-0.3" represents the results obtained by real dynamics for $\Delta t = 0.3$; "vir-0.3" stands for those produced by virtual dynamics for $\Delta t = 0.3$; *etc*. Statistical error bars are included.

**Fig. 6**. (Color online) Results for the average potential and the thermal fluctuation of the potential using different collision frequencies $\nu$ (unit: $fs^{-1}$) in the Andersen thermostat. Panels (a)-(b) for the $H_2O$ molecule. Three time intervals $\Delta t = 1.9$, 2.2, 2.4 (unit: fs) are used. Panels (c)-(d) for the $(Ne)_{13}$ cluster. Two time intervals $\Delta t = 20$, 50 (unit: fs) are used. [Panels (a) and (c) display the average potential energy per atom $\langle U(x)\rangle/\left(N_{atom}k_B\right)$ (unit: Kelvin). Panels (b) and (d) present the thermal fluctuation of the potential per atom $\sqrt{\langle U^2\rangle-\langle U\rangle^2}\Big/\left(N_{atom}k_BT\right)$.] "real-1.9" stands for the results obtained by real dynamics for $\Delta t = 1.9$ fs; "vir-1.9" represents those produced by virtual dynamics for $\Delta t = 1.9$ fs; *etc*.

Statistical error bars are included. [For comparison the converged results are obtained from our previous work on Langevin dynamics[10], with the friction coefficient ($\gamma$): Panels (a)-(b): $\gamma =$ 0.68 fs$^{-1}$, $\Delta t = 0.24$ fs for the $H_2O$ molecule; Panels (c)-(d): $\gamma = 0.001$ fs$^{-1}$, $\Delta t = 10$ fs for the $(Ne)_{13}$ cluster.]

**Fig. 7**. (Color online) Characteristic correlation time of the potential [panel (a)] and that of the Hamiltonian [panel (b)] for the Andersen thermostat for the quartic system $U(x) = x^4/4$. Three time intervals $\Delta t = 0.1$, 0.3, and 0.4 are used. All the parameters are in atomic units (a.u.). Statistical error bars are included. Hollow symbols with dashed lines: numerical results for the virtual dynamics case. Solid symbols: numerical results for the real dynamics case. "vir-0.1" represents the results for the virtual dynamics case for $\Delta t = 0.1$; "real-0.1" stands for the results for the real dynamics case for $\Delta t = 0.1$; *etc*.

**Fig. 8**. (Color online) Characteristic correlation time of the potential [panel (a)] and that of the Hamiltonian [panel (b)] for the Andersen thermostat for the $H_2O$ molecule at 100 K. Panels (c) and (d) are the same as panels (a) and (b), but for the $(Ne)_{13}$ cluster at 14 K. Two time intervals $\Delta t = 0.24$, 1.2 (unit: fs) are used for the $H_2O$ molecule in panels (a)-(b), while those used for the $(Ne)_{13}$ cluster at 14 K are $\Delta t = 20$, 50 (unit: fs). The unit of all the parameters is per femtosecond (fs$^{-1}$). Statistical error bars are included. Hollow symbols with dashed lines: numerical results for the virtual dynamics case. Solid symbols: numerical results for the real dynamics case. "vir-1.2" represents the numerical results for the virtual dynamics case for $\Delta t = 1.2$ fs; "real-1.2" stands for the real dynamics case for $\Delta t = 1.2$ fs; *etc*.

**Fig. 9**. (Color online) Comparison between Langevin dynamics and the Andersen thermostat on the accuracy of the results for the average potential energy and the thermal fluctuation of the potential of the 1-dimensional quartic potential $U(x)=x^4/4$ at $\beta=1$. The time interval $\Delta t=0.4$ is used. All the parameters are in atomic units (a.u.). Here $\xi$ stands for $\sqrt{2}\gamma$ for Langevin dynamics and $\nu$ for the Andersen thermostat. Statistical error bars are included. "Langevin real-0.4" represents the numerical results for the real dynamics case of Langevin dynamics for $\Delta t=0.4$; "Andersen vir-0.4" stands for the numerical results for the virtual dynamics case of the Andersen thermostat for $\Delta t=0.4$; *etc*.

**Fig. 10**. (Color online) Results for the average potential energy and the thermal fluctuation of the potential using different thermostat parameters ($\xi$) [friction coefficients $\sqrt{2}\gamma$ (unit: $fs^{-1}$) for Langevin dynamics; collision frequencies $\nu$ (unit: $fs^{-1}$) for the Andersen thermostat]. Panels (a)-(b) for the $H_2O$ molecule, the same time interval $\Delta t=2.4$ fs is used for Langevin dynamics and for the Andersen thermostat. Panels (c)-(d) for the $(Ne)_{13}$ cluster, the same time interval $\Delta t=50$ fs is used for the two types of thermostats. [Panels (a) and (c) display the average potential energy per atom $\langle U(x)\rangle/(N_{atom}k_B)$ (unit: Kelvin). Panels (b) and (d) present the thermal fluctuation of the potential per atom $\sqrt{\langle U^2\rangle-\langle U\rangle^2}\big/(N_{atom}k_BT)$.] Hollow symbols with dashed lines: numerical results for the virtual dynamics case. Solid symbols: numerical results for the real dynamics case. "Langevin real-2.4" represents the numerical results obtained by Langevin real dynamics for $\Delta t=2.4$ fs; "Andersen vir-2.4" stands for those produced by Andersen virtual dynamics for $\Delta t=2.4$ fs; *etc*.

**Fig. 11**. (Color online) For the quartic system $U(x)=x^4/4$. Panel (a) compares the characteristic correlation time of the potential produced by Langevin dynamics to that yielded by the Andersen thermostat for $\Delta t=0.1$, while Panel (c) does so for $\Delta t=0.4$. Panels (b) and (d) are the same as panels (a) and (c), respectively, but for the characteristic correlation

time of the Hamiltonian. All the parameters are in atomic units (a.u.). Here $\xi$ stands for $\sqrt{2}\gamma$ for Langevin dynamics and $\nu$ for the Andersen thermostat. Statistical error bars are included. Hollow symbols with dashed lines: numerical results for the virtual dynamics case. Solid symbols: numerical results for the real dynamics case. "Langevin real-0.1" represents the numerical results for the real dynamics case of Langevin dynamics for $\Delta t = 0.1$; "Andersen vir-0.4" stands for the numerical results for the virtual dynamics case of the Andersen thermostat for $\Delta t = 0.4$; *etc*.

**Fig. 12**. (Color online) Characteristic correlation time of the potential [Panel (a)] and that of the Hamiltonian [Panel (b)] for Langevin dynamics and for the Andersen thermostat for the $H_2O$ molecule at 100 K for the time interval $\Delta t = 0.24$ fs. Panels (c) and (d) are the same as Panels (a) and (b), respectively, but for the time interval $\Delta t = 2.4$ fs. The unit of all the thermostat parameters ($\xi$) is per femtosecond ($fs^{-1}$) for either of the Andersen collision frequency ($\nu$) and $\sqrt{2}$ times of the Langevin friction coefficient ($\sqrt{2}\gamma$). Statistical error bars are included. Hollow symbols with dashed lines: numerical results for the virtual dynamics case. Solid symbols: numerical results for the real dynamics case. "Langevin real-0.24" represents the numerical results obtained by Langevin real dynamics for $\Delta t = 0.24$ fs; "Andersen vir-0.24" stands for those produced by Andersen virtual dynamics for $\Delta t = 0.24$ fs; *etc*.

**Fig. 13**. (Color online) Same as Panels (a) and (b) of Fig. 12, but for the $(Ne)_{13}$ cluster at 14 K for the time interval $\Delta t = 50$ fs.

**Fig. 14.** (Color online) (a) For Langevin dynamics for the 1-dimensional harmonic system [Eq. (23)]. Unstable region is shown in grey. Stable region includes both real dynamics

and virtual dynamics cases. While the virtual dynamics case is shown in light cyan, the real dynamics case is separated into the underdamped part (in white) and the overdamped one (in light flavogreen) based on the Fourier transform of the position-displacement autocorrelation function (See Appendix C) in the sampling procedure. The optimal friction coefficient for configurational sampling satisfies the blue curve, while that for phase space sampling is depicted by the red curve. The former produces the minimum value of the characteristic time of the potential, while the latter leads to the minimum value of the characteristic time of the Hamiltonian. (b) Same as (a), but for the Andersen thermostat. The optimal collision frequency for configurational sampling satisfies the blue curve, while that for phase space sampling is described by the red curve.

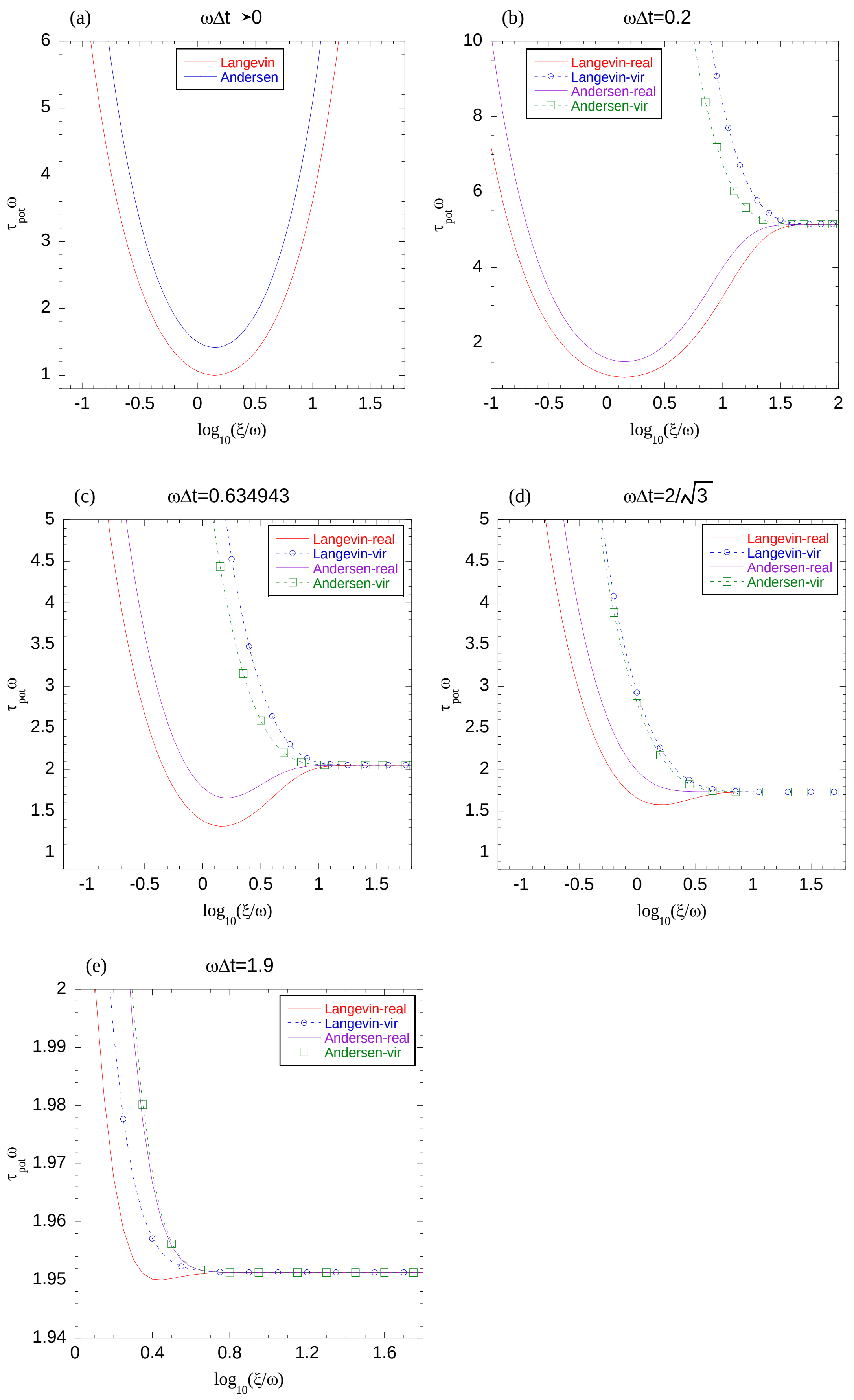

**Fig. 1**

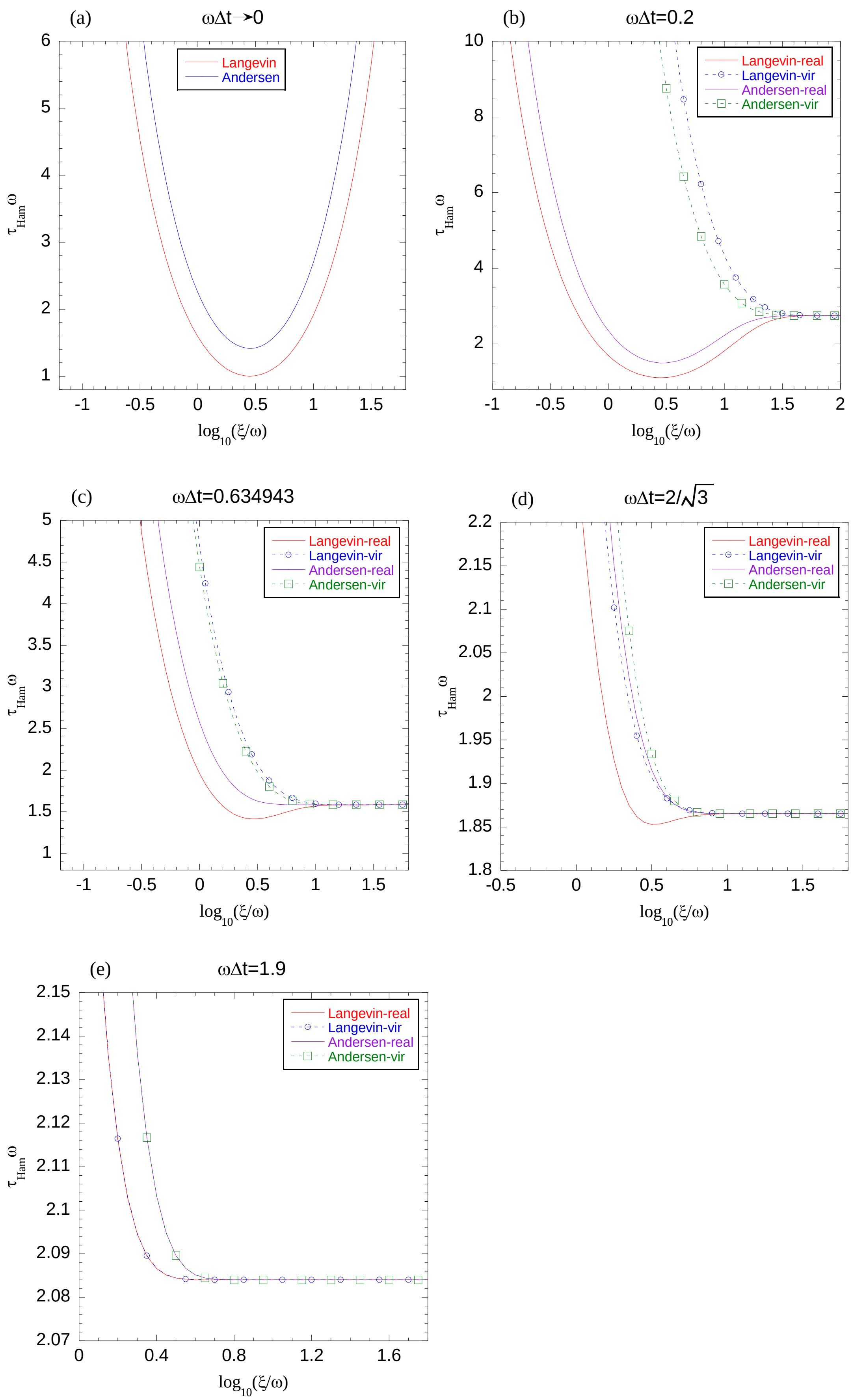

**Fig. 2**

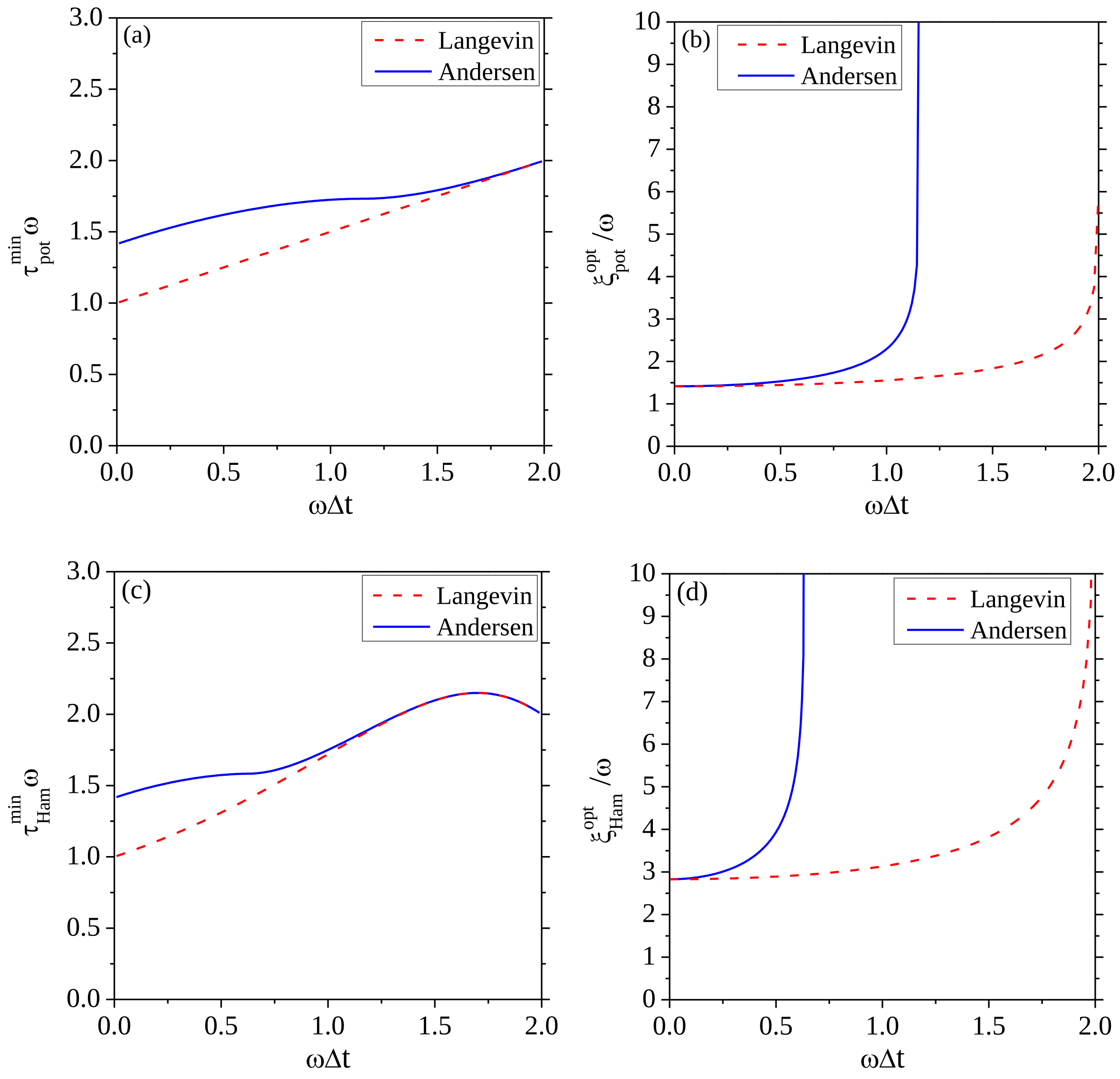

**Fig. 3**

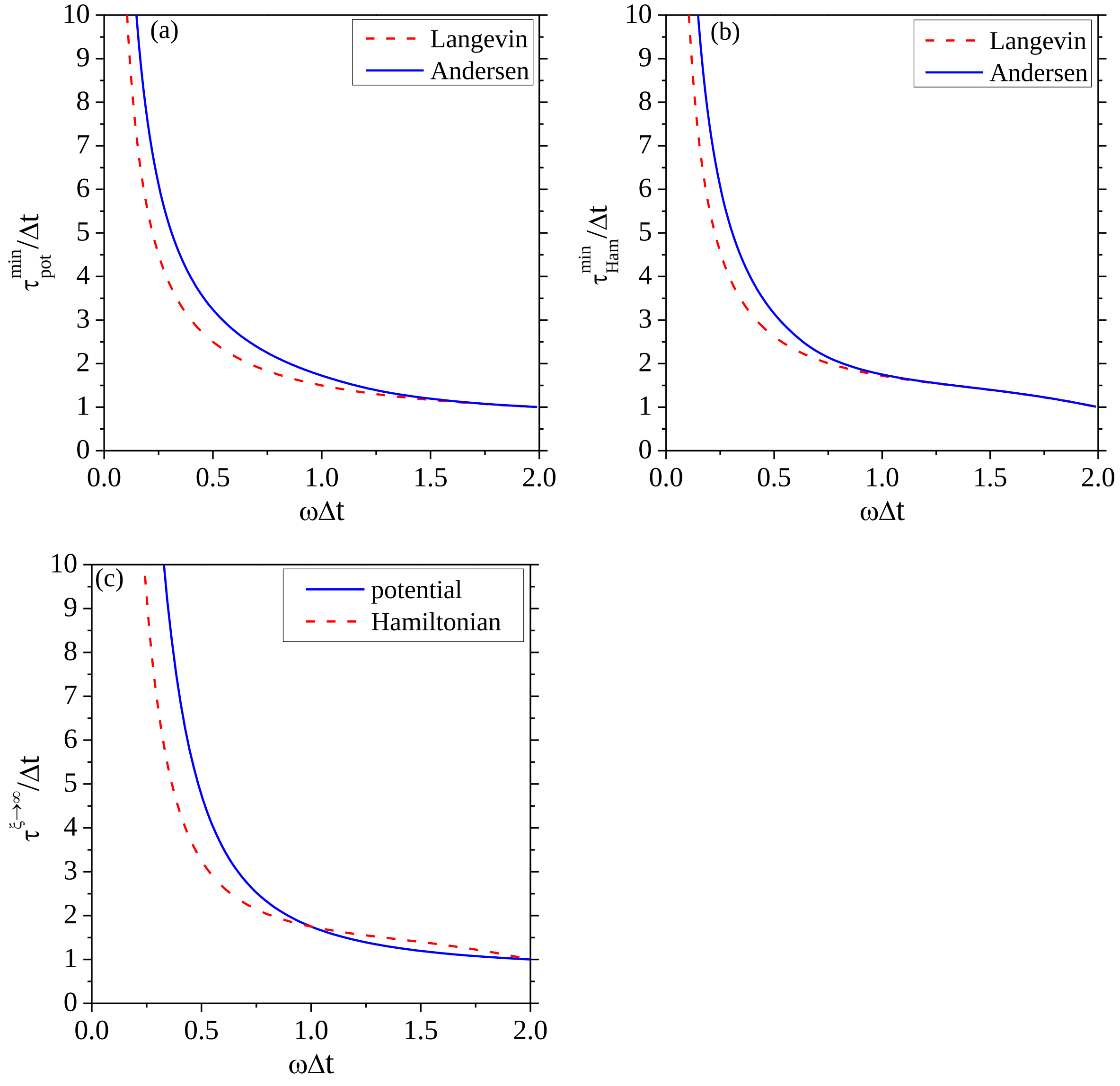

**Fig. 4**

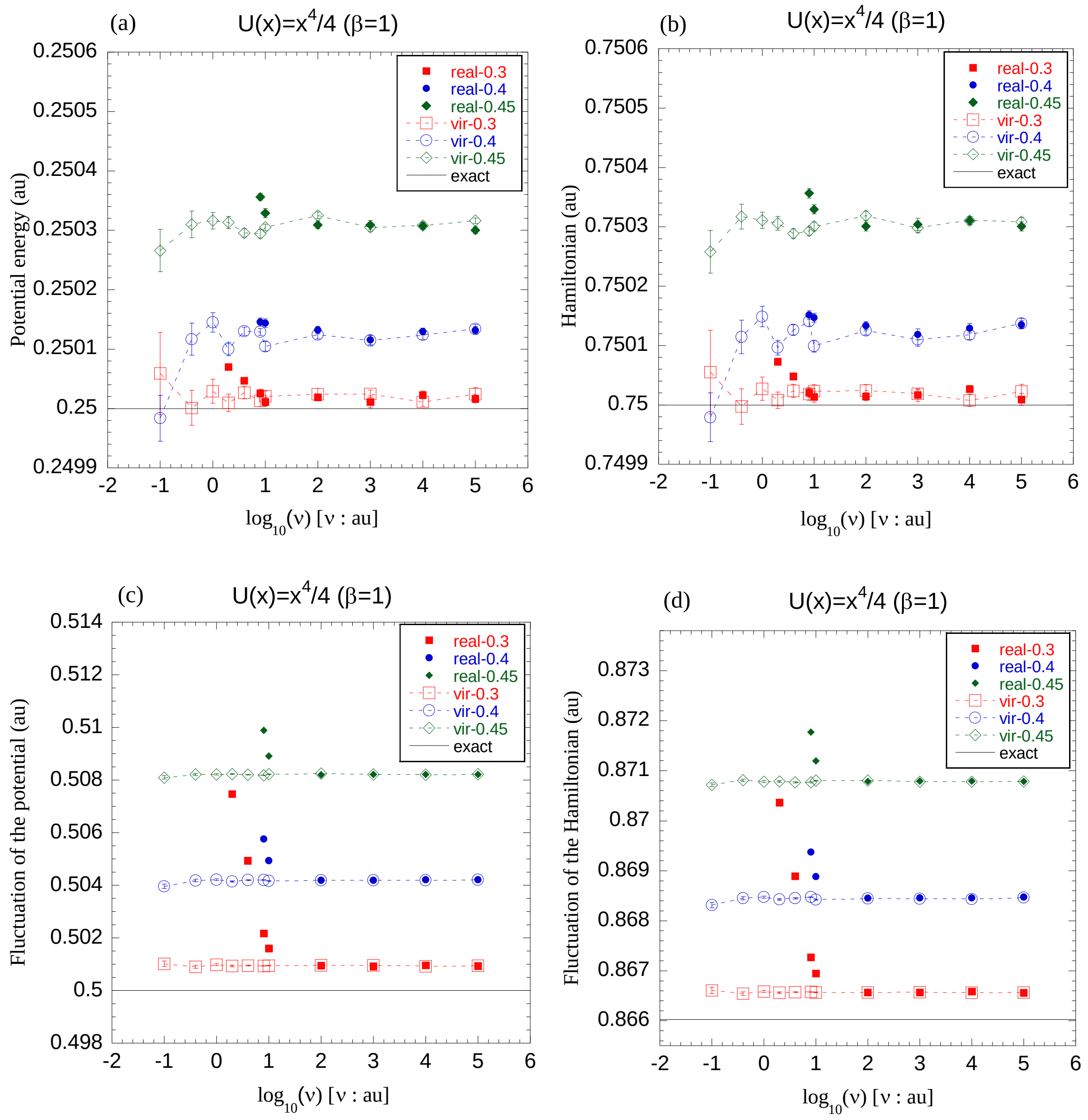

**Fig. 5**

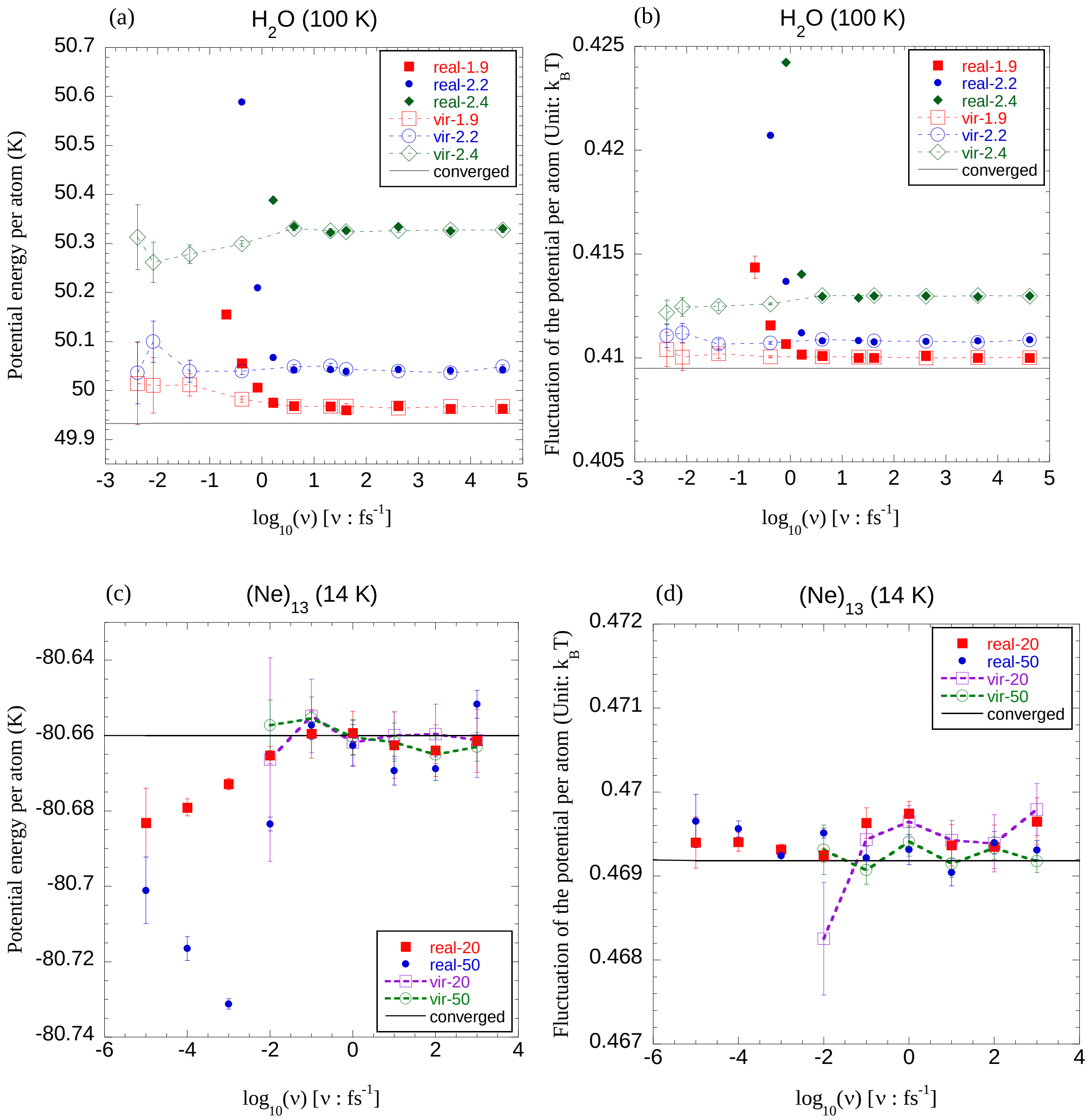

(a)
H2O (100 K)
Potential energy per atom (K)
log10(ν) [ν : fs-1]
real-1.9
real-2.2
real-2.4
vir-1.9
vir-2.2
vir-2.4
converged
(b)
H2O (100 K)
Fluctuation of the potential per atom (Unit: kBT)
log10(ν) [ν : fs-1]
(c)
(Ne)13 (14 K)
Potential energy per atom (K)
log10(ν) [ν : fs-1]
real-20
real-50
vir-20
vir-50
converged
(d)
(Ne)13 (14 K)
Fluctuation of the potential per atom (Unit: kBT)
log10(ν) [ν : fs-1]

**Fig. 6**

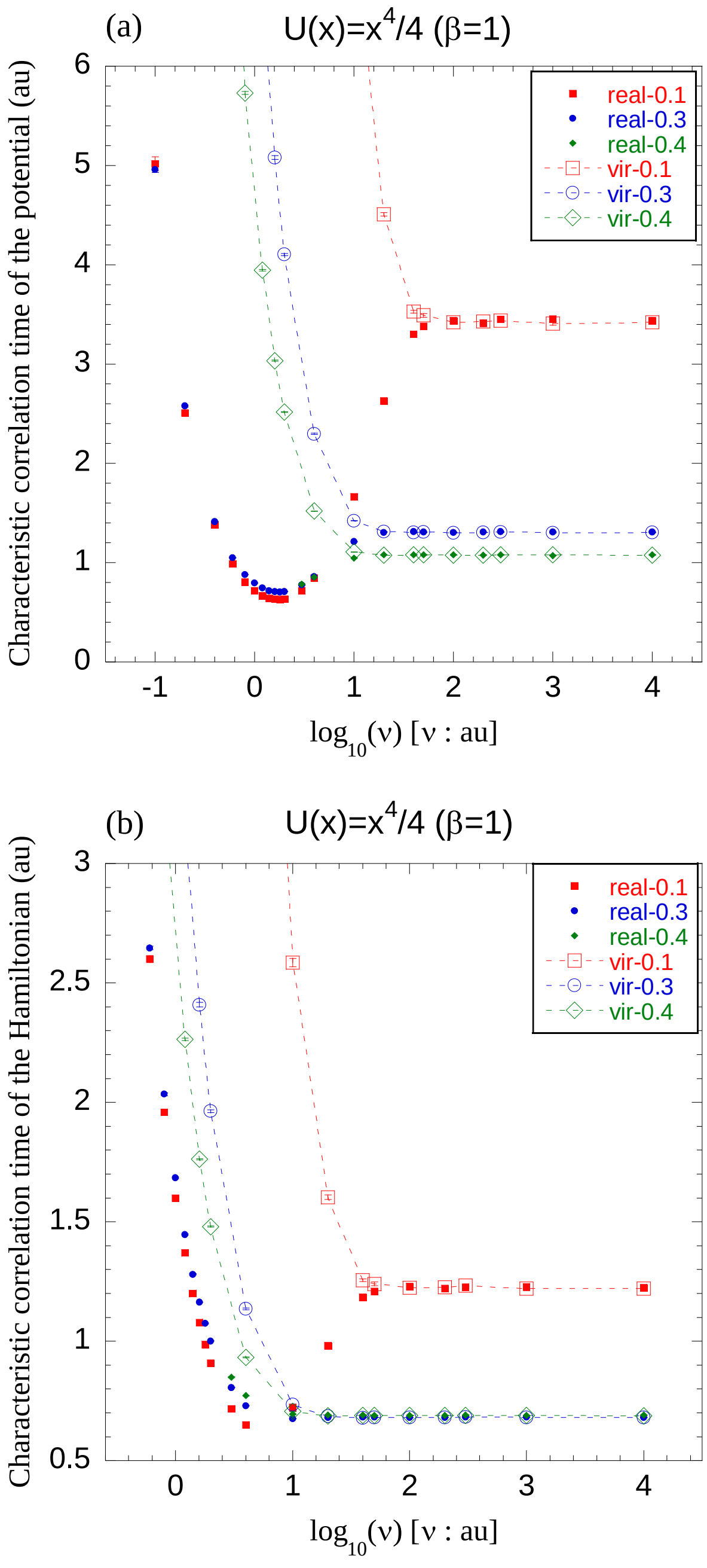

**Fig. 7**

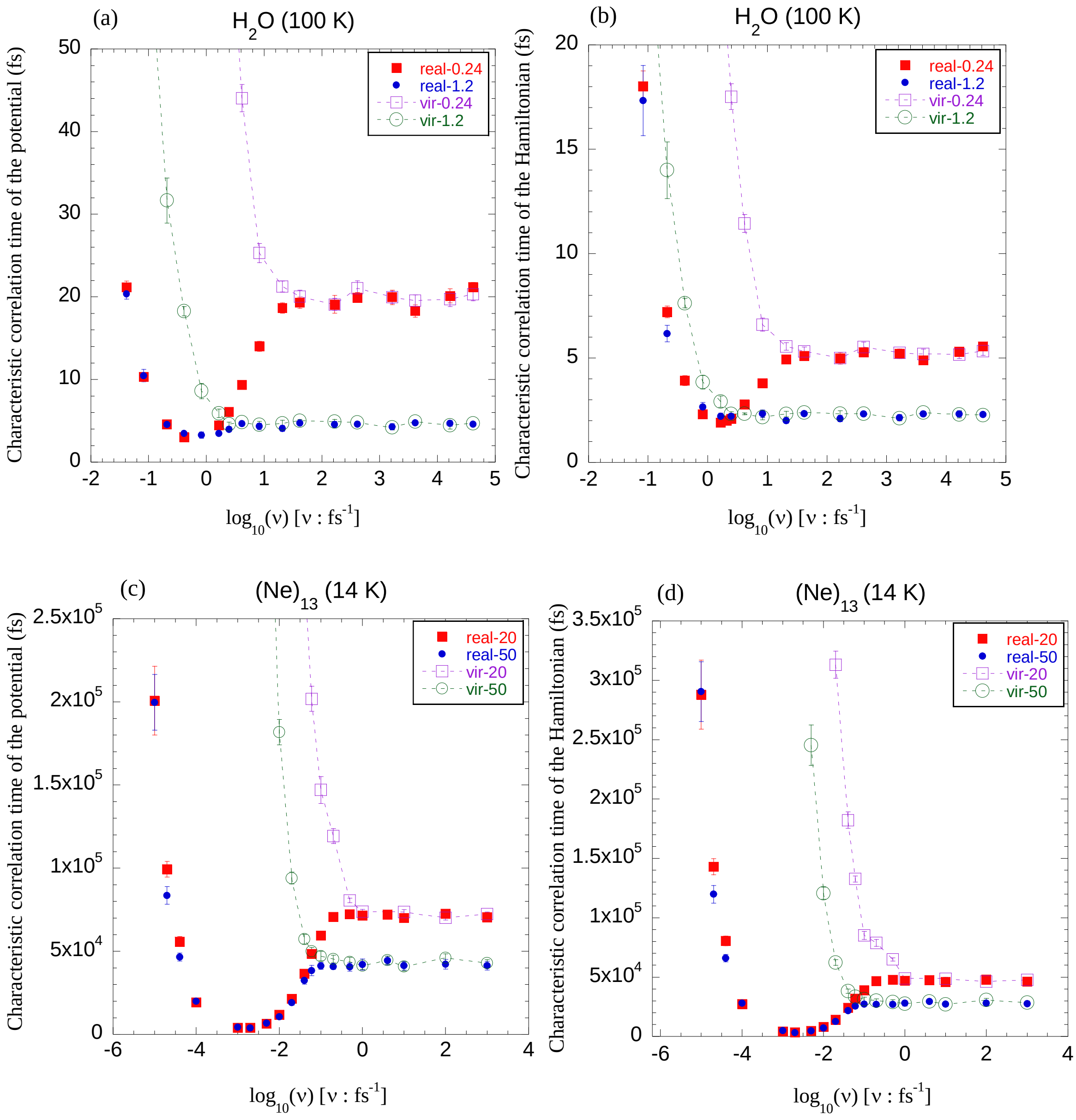
(a)
H2O (100 K)
Characteristic correlation time of the potential (fs)
real-0.24
real-1.2
vir-0.24
vir-1.2
log10(ν) [ν : fs-1]
(b)
H2O (100 K)
Characteristic correlation time of the Hamiltonian (fs)
real-0.24
real-1.2
vir-0.24
vir-1.2
log10(ν) [ν : fs-1]
(c)
(Ne)13 (14 K)
Characteristic correlation time of the potential (fs)
real-20
real-50
vir-20
vir-50
log10(ν) [ν : fs-1]
(d)
(Ne)13 (14 K)
Characteristic correlation time of the Hamiltonian (fs)
real-20
real-50
vir-20
vir-50
log10(ν) [ν : fs-1]

**Fig. 8**

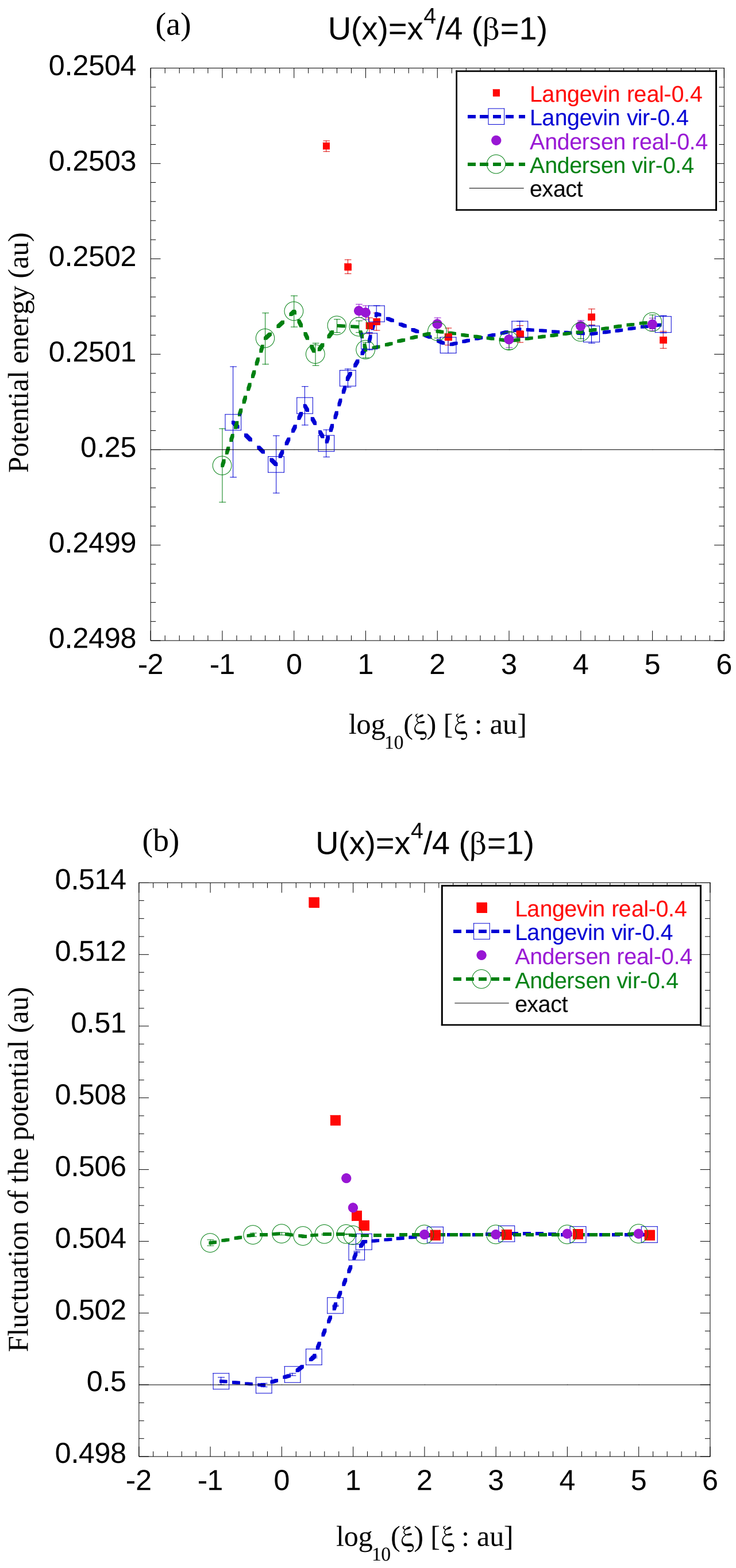

**Fig. 9**

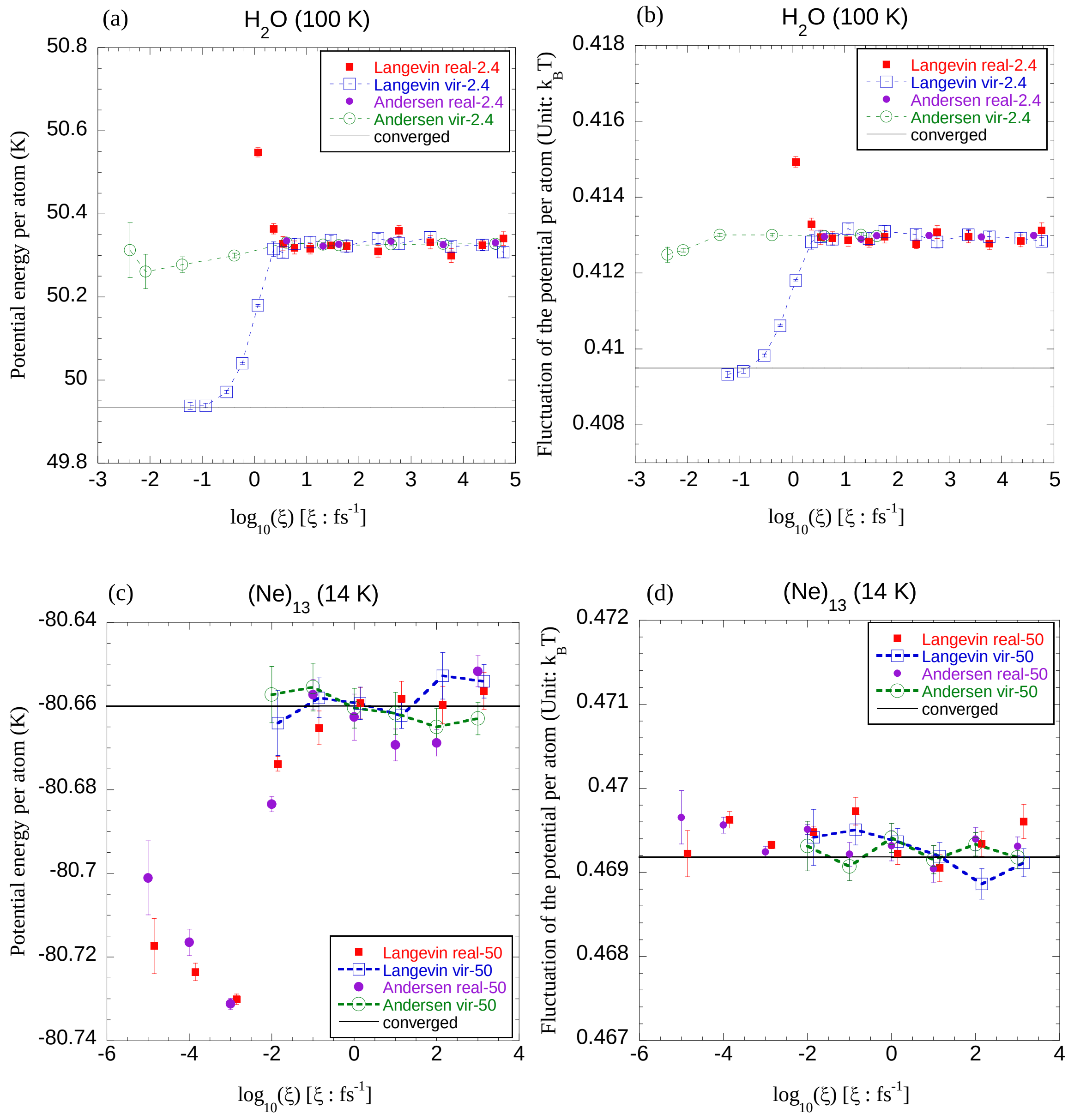

(a)
H2O (100 K)
Potential energy per atom (K)
log10(ξ) [ξ : fs-1]
Langevin real-2.4
Langevin vir-2.4
Andersen real-2.4
Andersen vir-2.4
converged
(b)
H2O (100 K)
Fluctuation of the potential per atom (Unit: kBT)
(c)
(Ne)13 (14 K)
Langevin real-50
Langevin vir-50
Andersen real-50
Andersen vir-50
(d)
(Ne)13 (14 K)

**Fig. 10**

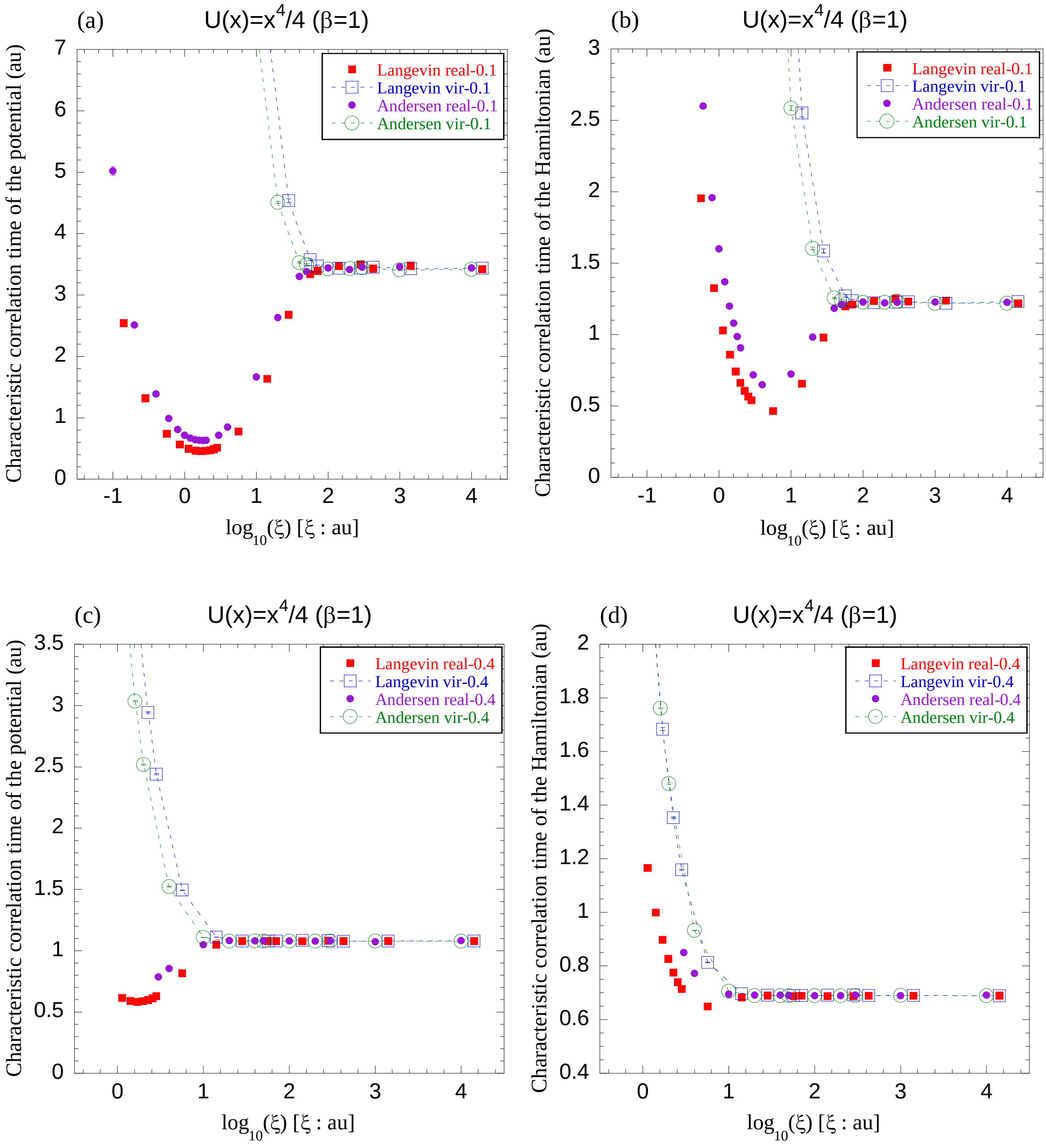

**Fig. 11**

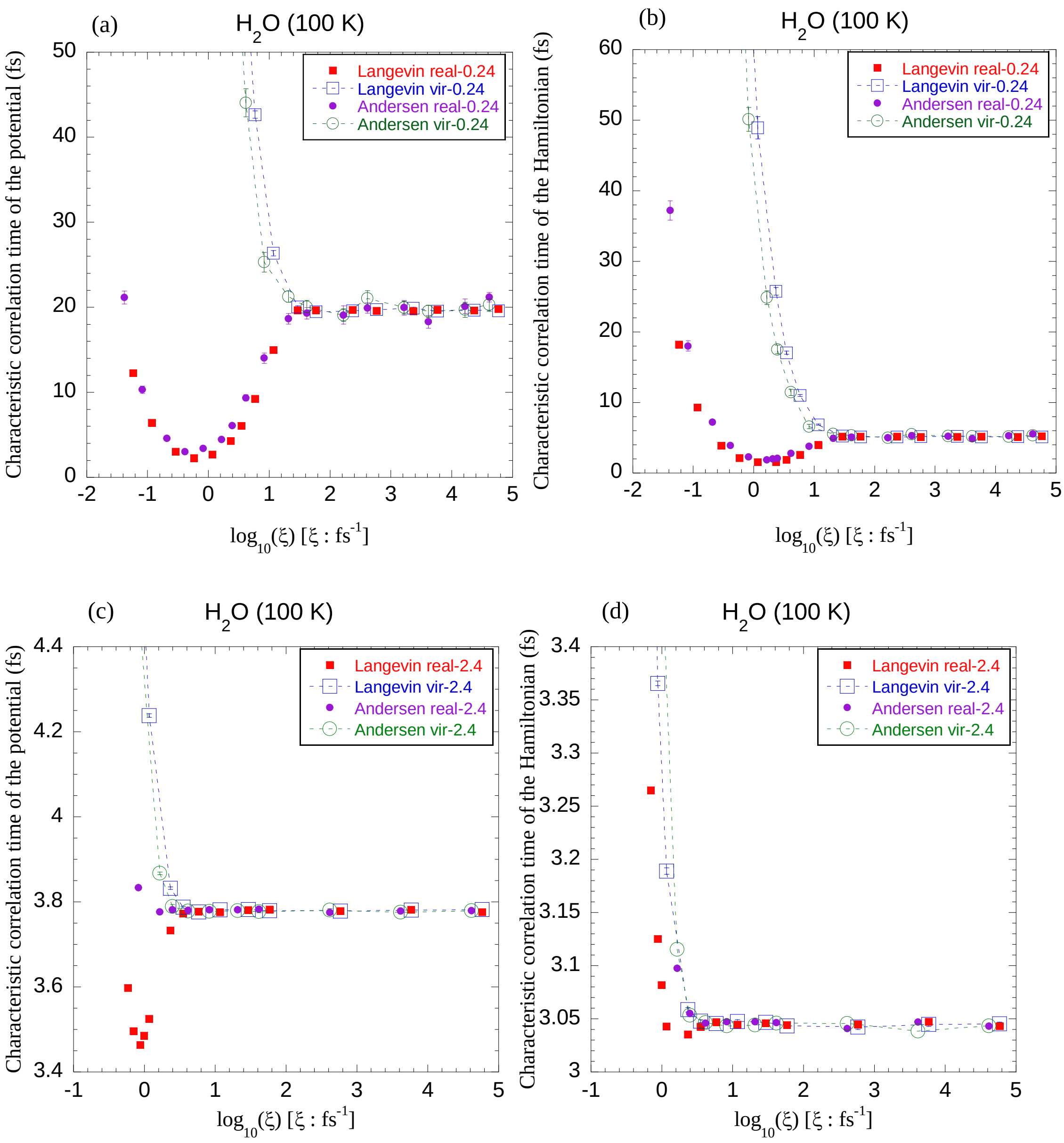
(a)
H2O (100 K)
Characteristic correlation time of the potential (fs)
log10(ξ) [ξ : fs-1]
Langevin real-0.24
Langevin vir-0.24
Andersen real-0.24
Andersen vir-0.24
(b)
H2O (100 K)
Characteristic correlation time of the Hamiltonian (fs)
log10(ξ) [ξ : fs-1]
Langevin real-0.24
Langevin vir-0.24
Andersen real-0.24
Andersen vir-0.24
(c)
H2O (100 K)
Characteristic correlation time of the potential (fs)
log10(ξ) [ξ : fs-1]
Langevin real-2.4
Langevin vir-2.4
Andersen real-2.4
Andersen vir-2.4
(d)
H2O (100 K)
Characteristic correlation time of the Hamiltonian (fs)
log10(ξ) [ξ : fs-1]
Langevin real-2.4
Langevin vir-2.4
Andersen real-2.4
Andersen vir-2.4

**Fig. 12**

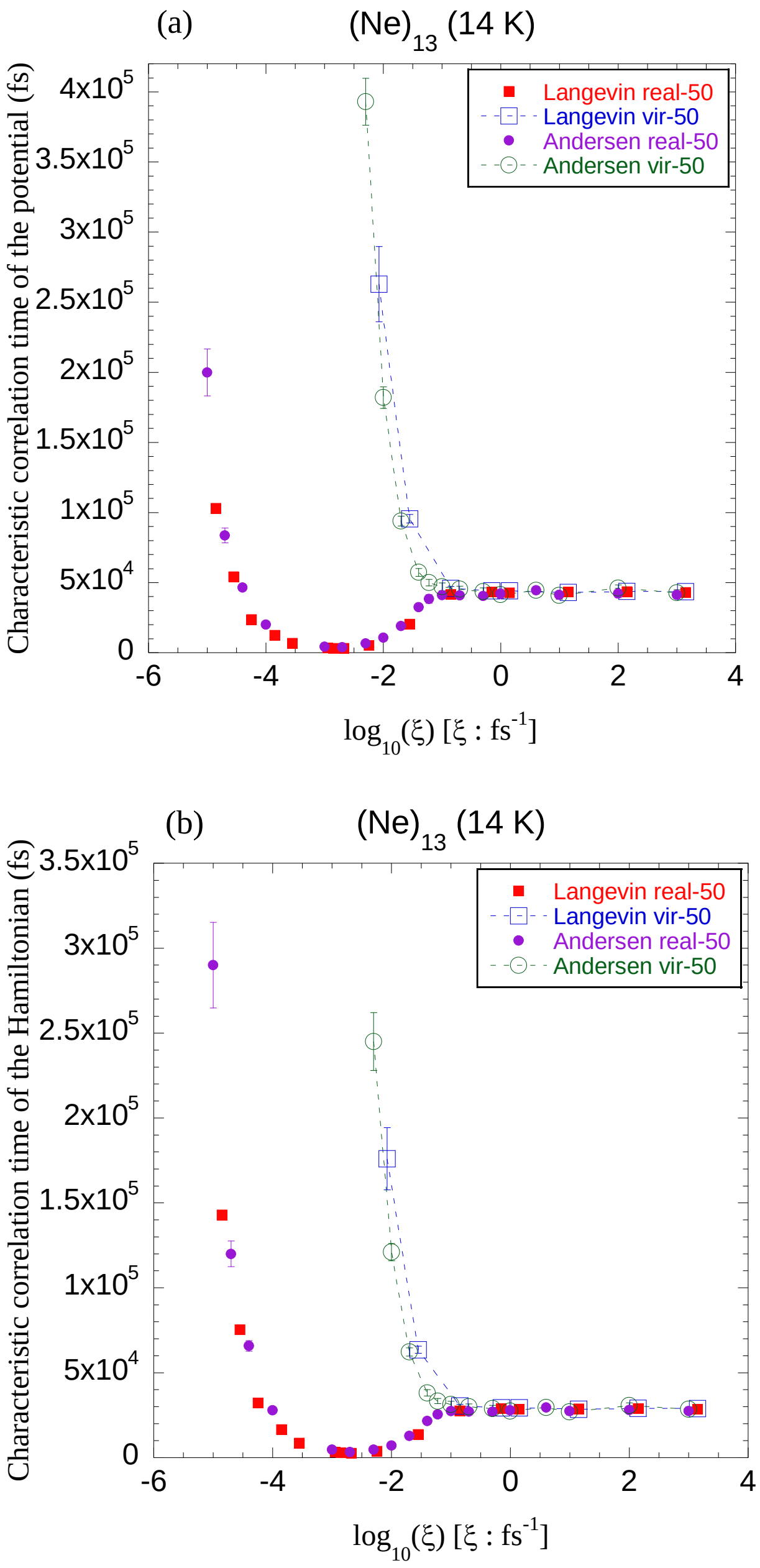
(a)
(Ne)13 (14 K)
Characteristic correlation time of the potential (fs)
Langevin real-50
Langevin vir-50
Andersen real-50
Andersen vir-50
log10(ξ) [ξ : fs-1]
(b)
(Ne)13 (14 K)
Characteristic correlation time of the Hamiltonian (fs)
Langevin real-50
Langevin vir-50
Andersen real-50
Andersen vir-50
log10(ξ) [ξ : fs-1]

**Fig. 13**

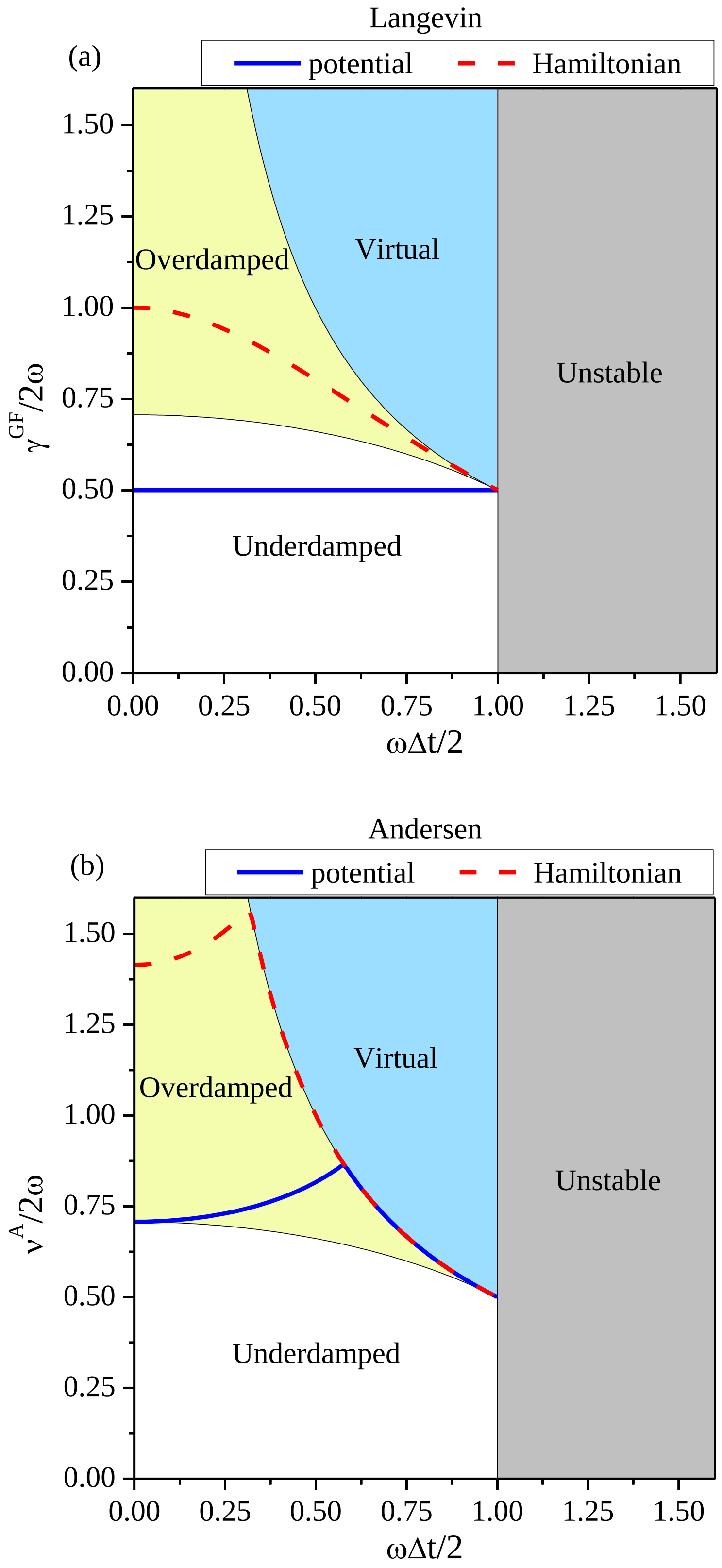

**Fig. 14**